\documentclass[epj]{svjour}
\usepackage{graphics}
\begin{document}
\title{Microscopic Calculation of the Constitutive Relations}
\author{Christian Brouder\inst{1}, St\'ephanie Rossano\inst{2}}
\institute{Laboratoire de Min\'eralogie-Cristallographie,
   CNRS UMR7590, Universit\'es Paris 6 et 7, IPGP, 4 place Jussieu,
  75252 Paris Cedex 05,
  France. {\small \tt brouder@lmcp.jussieu.fr} \and
  Laboratoire des G\'eomat\'eriaux, IFG,
  Universit\'e de Marne la Vall\'ee,
  Cit\'e Descartes - Champs-sur-Marne
 77454 Marne La Vall\'ee cedex 2, France
{\small \tt rossano@univ-mlv.fr}}
\date{\today}
\abstract{
Homogenization theory is used to calculate the macroscopic dielectric constant
from the quantum microscopic dielectric function in a periodic medium. The method
can be used to calculate any macroscopic constitutive relation, but it
is illustrated here for the case of electrodynamics of matter. The so-called
cell problem of homogenization theory is solved and an explicit expression
is given for the macroscopic dielectric constant in a form akin to the
Clausius-Mossotti or Lorentz-Lorenz relation.
The validity of this expression is checked by showing that the macroscopic
dielectric constant is causal and has the expected symmetry properties,
and that the average of the microscopic energy density is the
macroscopic one. Finally, the general expression is applied to Bloch
eigenstates. Finally, the corresponding many-body problem is
briefly discussed.
\PACS{
      {77.22.Ch}{Permittivity (dielectric function)}   \and
      {78.20.Ci}{Optical constants}   \and
      {41.20.Bt}{Maxwell equations}
     } 
} 

\maketitle

\section{Introduction}

When light falls onto a crystal, the quantum interaction of
light with matter is represented locally by
a microscopic dielectric function
$\epsilon({\bf{r}},{\bf{r}}')$. 
To calculate the macroscopic (homogeneous) dielectric
constant, one considers that
the charge distribution created by the light polarizes
the crystal which, in turn, reacts by inducing 
an electric field that modifies
the charge distribution.

In quantum chemistry, this reaction field
represents the influence of the solvent on the
solute \cite{Ruiz,Rivail,Tawa}.
Other descriptions use the related concept of
local fields \cite{Schnatterly,Keller}, which
has been described, for disordered media, 
by a cluster expansion
\cite{Felderhof,Cichocki,Hinsen,Ersfeld}.

For periodic media, the local field effect
was evaluated in the early sixties by
Adler \cite{Adler} and Wiser \cite{Wiser}.
However, 
many textbooks in solid-state physics still identify the 
macroscopic dielectric constant 
$\bar\epsilon$ with $\langle \epsilon\rangle $, the average of
the microscopic dielectric function 
$\epsilon({\bf{r}},{\bf{r}}')$ over a unit cell.
Even the most cautious authors \cite{Kranendonk,Barron,Henderson}
do not go beyond the relation
$\bar\epsilon=\epsilon_0(\epsilon_0+2\langle \epsilon\rangle)/
(4\epsilon_0-\langle \epsilon\rangle)$,
that goes by the name of Clausius-Mossotti or Lorentz-Lorenz
(see Refs.\cite{Scaife,Lyubimov1,Lyubimov2} for a history of this
relation). Since the local field effect can be quite large
\cite{Vechten}, its neglect can probably be attributed to
the numerical burden of the standard local field formula
\cite{Adler,Wiser}. 

In this paper, homogenization
theory will be used to provide various alternative formulas
for the calculation of the local field effects.
Homogenization theory is tailored to 
calculate the macroscopic $\bar\epsilon$ from
the microscopic $\epsilon({\bf{r}},{\bf{r}}')$.
The major trick of the method is to expand all fields
as a series in ascending powers of the ratio of the
lattice parameter over the wavelength of the external
electromagnetic field. Physicists have sometimes used
such an expansion \cite{Mochan2}, but mathematicians
exploited it extensively and turned it into a rigorous 
tool.
Homogenization theory is a branch of applied mathematics
that started its expansion in the late seventies 
\cite{Bensoussan,Sanchez} 
to understand the macroscopic properties of 
composite, porous, disordered, bubbly, fibrous or layered 
materials. It is now a fully fledged theory \cite{Jikov}
that has been applied successfully in
many areas, such as mechanics, acoustics, electrostatics,
fluid dynamics, statistical physics, numerical analysis,
materials sciences, electromagnetism \cite{Maso},
petroleum geophysics \cite{Yongji},
shape-memory alloys \cite{Bhattacharya}
and pile foundation analysis \cite{Aganovic}.

This paper starts with an introduction to homogenization theory,
then the microscopic Maxwell equations are homogenized to yield
the macroscopic Maxwell equations and the constitutive relations.
Since homogenization theory is not a usual tool of solid state 
physics, the calculations in these sections will be given in detail.
The rest of the article will follow the elliptic style of normal
research papers.  Various formulas will be given, corresponding
to different physical situations. Then, several desirable 
properties of the dielectric constant will be derived. It
will be shown that the macroscopic dielectric
constant is causal, has the proper symmetry properties and gives
the expected energy density and Poynting vector. The
formalism is adapted to the case of Bloch eigenstates
and a band-structure formula for $\bar\epsilon$ is given.
Finally, the much simpler case of the many-body 
dielectric function is briefly discussed.

\section{The point of view of homogenization theory}
Ever since the nineteenth century, physicists have homogenized microscopic
systems by performing averages over distances very small compared to
the macroscopic wavelength and very large compared to the atomic dimensions.
Mathematical homogenization started when homogenization was not
considered any longer as an averaging operation, but as a limit process.
As an example, consider that a microscopic quantity can be represented
by the periodic function $f(x)=\sin x +b$. To express the fact that the
oscillation is very fast, mathematicians tried to give a meaning to the limit
of $f(x/a)$ as $a\rightarrow 0$. In other words, what is the limit of a
periodic function when its period becomes infinitely small ?

\begin{figure}
\resizebox{0.45\textwidth}{!}{%
  \includegraphics{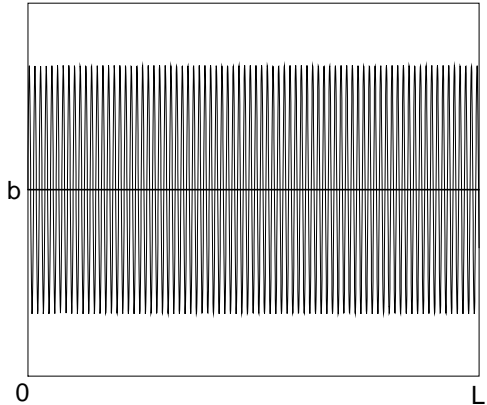}
}
\caption{Limit of $\sin(x/a)+b$ as $a\rightarrow 0$}
\label{fig:1}       
\end{figure}

The contact with the physical point of view comes from the fact that
the limit is the average of $f(x)$ over a period : $f(x/a)\rightarrow b$
as $a\rightarrow 0$. However, figure 1 shows that this limit is somewhat
unusual. Physicists generally do not bother very much with the various 
mathematical limits, but in the present case, it is necessary to realize that
we do not deal with a strong limit, but with a weak limit. 
Strong convergence of $f(x)$ to its limit $b$, which means that,
for a fixed $L$,
$\int_0^L dx |f(x/a)-b|\rightarrow 0$ as $a\rightarrow 0$, is clearly 
not realized for oscillating functions
(for our example, $\int_0^L dx |f(x/a)-b|\rightarrow 2L/\pi$).
Oscillating functions enjoy only weak convergence,
which is defined by the fact that,
for any smooth function $g(x)$, 
$\int_0^L dx g(x)f(x/a)\rightarrow b\int_0^L dx g(x)$ as $a\rightarrow 0$.
In physical terms, the fast oscillations of $f(x/a)$ are smoothed away by
measuring $f(x/a)$ with a device having a finite resolution $g(x)$.

Strong convergence enjoys many nice properties, for instance the
product of two strongly convergent functions converges
to the product of the limits, but no such thing is
available for weak convergence. In our example,
the weak limit of $f^2(x/a)$ is $b^2+1/2$ and not $b^2$.
Since it is clear that the average of the product of two functions
is not the product of the averages of the functions, the reader
may wonder why it is useful to consider homogenization as
a limit instead of an average. The advantage of homogenization
as a limit is twofold. 
On the first hand, it leads naturally to an asymptotic expansion of the
functions in terms of $a$ 
(i.e. $f(x/a)\rightarrow b+a f^{(1)}(x/a)+\cdots$),
so that corrections to the average become available. On the other hand,
it enables us to treat not only functions but also
differential equations.

The way mathematical homogenization deals with differential
equations can be illustrated
by a simple example \cite{Bensoussan}. Consider a wire of length $L$,
uniformly charged with a constant charge density $n$.
A potential $V$ is applied between both ends of the
wire, and the wire is assumed to have a periodic
structure represented by the periodic dielectric function
$\epsilon(x/a)$. Standard electrostatic theory
tells us that there is a potential $\phi(x)$ such
that the electric field $e(x)=-\phi'(x)$. The 
microscopic constitutive relation is 
$d(x)=\epsilon(x/a)e(x)$, and the electrostatic
equation $d'(x)=n$ gives us the
equation $(\epsilon(x/a)\phi'(x))'=-n$,
with the boundary conditions $\phi(0)=0$, $\phi(L)=V$.
For simplicity, we assume that $L$ corresponds to an
integer number of periods $2\pi a$. This electrostatic
equation has a unique solution $\phi(x)$ for each $a$
(that can be obtained explicitly). From the explicit
solution, it can be shown that $\phi(x)$ tends weakly to a
function $\Phi(x)$ when $a$ tends to zero
(with respect to the wire length $L$).
The purpose
of homogenization theory is to answer the question:
to which differential equation is $\Phi(x)$ a solution?
Again, the explicit solution can be used to show
that $\Phi(x)$ is the solution of 
$(\bar\epsilon\Phi'(x))'=-n$,
with $\Phi(0)=0$, $\Phi(L)=V$ and $\bar\epsilon$
is a constant. The surprising fact is
that $\bar\epsilon$ is not given by the average of
$\epsilon(x)$ over a period, but by the inverse of
the average of $1/\epsilon(x)$ over a period.
However, such a simple relation between $\epsilon(x)$
and $\bar\epsilon$ is restricted to one-dimensional
problems, and more work is required in
three dimensions.

More generally, for a differential equation with
rapidly oscillating coefficients, homogenization theory
determines whether the solution has a limit when the
period tends to zero, and to which equation the limit
is a solution. This is exactly what we need to derive
constitutive relations from a microscopic description
of matter. Homogenization theory has shown that the
macroscopic equation can be strongly different from the
microscopic one. For instance, instantaneous microscopic
equations can turn into equations with memory
\cite{Sanchez3,Alexandre},
local equations can develop non-local macroscopic terms
\cite{Tartar}, mixtures of optically inactive
materials can become optically active 
\cite{Bossavit2,ElFeddi}

\section{Functional transformations}
In this section, some functional transformations are introduced.
 
For the purpose of homogenization, a function
$f({\bf{r}})$ is written as
${\sf{f}}({\bf{R}},{\bf{\rho}})$, where the dependence 
of ${\sf{f}}$ is slow for the variable ${\bf{R}}$ and
periodic for the variable ${\bf{\rho}}$.
This so-called two-scale correspondence can be done explicitly as follows.
Take a three-dimensional periodic lattice with
Brillouin zone
$BZ$. Let $C$ be the Wigner-Seitz cell of the lattice
and $|C|$ its volume. The crystal sites will be denoted by
${\bf{R}}_s$, and the reciprocal lattice vectors by
${\bf{K}}$.

Write the function $f$ as a Fourier transform
\begin{eqnarray*}
f({\bf{r}})&=&\int d{\bf{q}} \exp(i{\bf{q}}\cdot{\bf{r}})
  \tilde{f}({\bf{q}})\\
&=&\sum_{{\bf{K}}}\int_{BZ} d{\bf{q}}
  \exp[i({\bf{q}}+{\bf{K}})\cdot{\bf{r}}]
  \tilde{f}({\bf{q}}+{\bf{K}}).
\end{eqnarray*}
Define now
\begin{eqnarray*}
{\sf{f}}({\bf{R}},{\bf{\rho}})=\int_{BZ} d{\bf{q}}
  \exp(i{\bf{q}}\cdot{\bf{R}})\tilde{{\sf{f}}}({\bf{q}},{\bf{\rho}}),
\end{eqnarray*}
with 
\begin{eqnarray*}
  \tilde{{\sf{f}}}({\bf{q}},{\bf{\rho}})&=&\sum_{{\bf{K}}}
  \exp(i {\bf{K}}\cdot{\bf{\rho}})
  \tilde{f}({\bf{q}}+{\bf{K}}) \quad\mathrm{for}\quad 
  \mathbf{q} \in BZ\\
  \tilde{{\sf{f}}}({\bf{q}},{\bf{\rho}})&=&0
  \quad\mathrm{for}\quad \mathbf{q} \notin BZ
\end{eqnarray*}
Then it is clear that $f({\bf{r}})= {\sf{f}}({\bf{r}},{\bf{r}})$,
so that
\begin{eqnarray}
f(\mathbf{r})=\int_{BZ} d{\bf{q}}
  \exp(i{\bf{q}}\cdot{\bf{r}})\tilde{{\sf{f}}}({\bf{q}},{\bf{r}}).
\label{fder}
\end{eqnarray}

\begin{figure}
\resizebox{0.45\textwidth}{!}{%
  \includegraphics{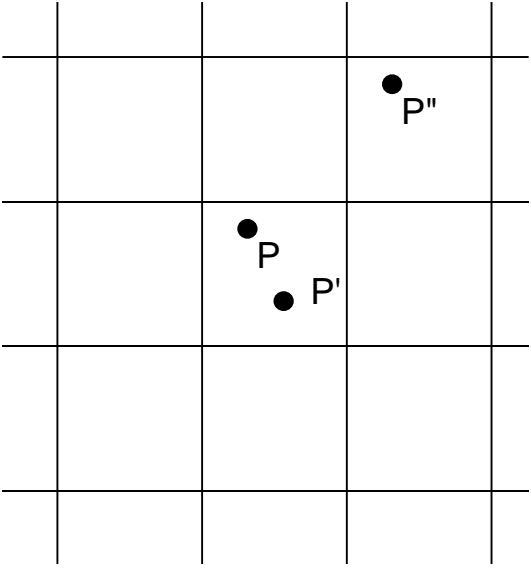}
}
\caption{$P$ and $P'$ are two points in the same
 cell, $P$ and $P''$ are separated by a lattice vector}
\label{fig:2}       
\end{figure}
${\sf{f}}({\bf{R}},{\bf{\rho}})$ has the lattice periodicity 
in ${\bf{\rho}}$ and depends more slowly on ${\bf{R}}$
than on ${\bf{\rho}}$. More precisely, homogenization
is useful when the function $f(\mathbf{r})$ 
varies slowly from $P$ to $P''$ (see Fig.2),
i.e. for two points
that differ by a (short) lattice vector and varies
arbitrarily from $P$ to $P'$ inside a cell. 
Then $f(\mathbf{r})$ is transformed into
${\sf{f}}({\bf{R}},{\bf{\rho}})$, the $\rho$ variable
describes the fast variation of the function inside the
cell, and the $\mathbf{R}$ variable its smooth variation from cell to cell.

This representation of a function of one variable by a
function of a fast periodic variable ${\bf{\rho}}$ and a slower 
variable ${\bf{R}}$ is the essence of the two scale analysis
of homogenization theory. 

We shall not homogenize the Maxwell equations with
the usual two-scale function ${\sf{f}}({\bf{R}},{\bf{\rho}})$,
but with its Fourier transform with respect to the slow 
variable $\mathbf{R}$:
$\tilde{\sf{f}}({\bf{q}},{\bf{\rho}})$. To establish
a direct link between $\tilde{\sf{f}}({\bf{q}},{\bf{\rho}})$
and $f({\bf{r}})$ and for future reference, we shall make use
of the standard relations for infinite Born-von K\'arm\'an
boundary conditions \cite{Calais} (which are derived from
the Poisson summation formula \cite{Duran}):
\begin{eqnarray}
\int_{BZ}d{\bf{q}}\exp(i{\bf{q}}\cdot{\bf{R}}_s)&=&
\frac{(2\pi)^3}{|C|}\delta_{s,0}\nonumber\\
\int_{C}d{\bf{\rho}}\exp(i{\bf{K}}\cdot{\bf{\rho}})&=&
|C|\delta_{{\bf{K}},0}\nonumber \\
\sum_{\bf{K}}\exp(i{\bf{K}}\cdot{\bf{\rho}})&=&
|C|\sum_s\delta({\bf{\rho}}+{\bf{R}}_s)\nonumber \\
\sum_s\exp(i{\bf{q}}\cdot{\bf{R}}_s)&=&
\frac{(2\pi)^3}{|C|}\sum_{\bf{K}}\delta({\bf{q}}-{\bf{K}}).
\label{somsurs}
\end{eqnarray}
An infinite Born-von K\'arm\'an boundary condition
avoids the subtle problems linked with the use of 
a finite Born-von K\'arm\'an domain, for instance
the question whether or not the other domains
contribute to the reaction field.

Using the definition of the Fourier transform we find
\begin{eqnarray}
\tilde{\sf{f}}({\bf{q}},{\bf{\rho}})&=&
  \sum_{{\bf{K}}} \exp(i{\bf{K}}\cdot{\bf{\rho}})
  \tilde{f}({\bf{q}}+{\bf{K}})\nonumber\\
&&\hspace*{-2mm}=\frac{1}{(2\pi)^3}
\sum_{{\bf{K}}} \exp(i{\bf{K}}\cdot{\bf{\rho}})
  \int d{\bf{r}}\exp[-i({\bf{q}}+{\bf{K}})\cdot{\bf{r}}]
  f({\bf{r}})\nonumber \\
&&\hspace*{-2mm}=\frac{1}{(2\pi)^3}  \int d{\bf{r}} f({\bf{r}})
\exp(-i{\bf{q}}\cdot{\bf{r}})
\sum_{{\bf{K}}} \exp[i{\bf{K}}\cdot({\bf{\rho}}-{\bf{r}})]
\nonumber \\
&&\hspace*{-2mm}=\frac{|C|}{(2\pi)^3} \sum_s 
  \int d{\bf{r}} f({\bf{r}})
\exp(-i{\bf{q}}\cdot{\bf{r}})
\delta({\bf{\rho}}-{\bf{r}}+{\bf{R}}_s) \nonumber \\
&&\hspace*{-2mm}=\frac{|C|}{(2\pi)^3} \sum_s 
f({\bf{\rho}}+{\bf{R}}_s)
\exp[-i{\bf{q}}\cdot({\bf{\rho}}+{\bf{R}}_s)].
\label{deraqrho}
\end{eqnarray}

For a function $f(\mathbf{r})$, with
two-scale transform ${\sf{f}}({\bf{R}},{\bf{\rho}})$,
we define the average over a unit cell by
\begin{eqnarray*}
\langle f({\bf{r}})\rangle &=&
\langle {\sf{f}}({\bf{R}},{\bf{\rho}})\rangle =
\frac{1}{|C|}\int_C d{\bf{\rho}}\,{\sf{f}}({\bf{R}},{\bf{\rho}})\\
&=&
 \int_{BZ} d{\bf{q}} \tilde{f}({\bf{q}})\exp(i{\bf{q}}\cdot{\bf{R}}).
\end{eqnarray*}
Similarly, for $\tilde{\sf{f}}({\bf{q}},{\bf{\rho}})$
\begin{eqnarray*}
\langle \tilde{\sf{f}}({\bf{q}},{\bf{\rho}})\rangle &=&
\frac{1}{|C|}\int_C d{\bf{\rho}}\tilde{\sf{f}}({\bf{q}},{\bf{\rho}})\\
&=&
 \tilde{f}({\bf{q}}).
\end{eqnarray*}

Since ${\bf{q}}$ is in the first Brillouin zone, the average 
over a unit cell has the effect of stripping all the high frequency
components off the Fourier transform of $f$.
We shall use this average to go from microscopic to
macroscopic fields. The macroscopic fields
will be microscopic fields averaged over a unit cell,
so that all Fourier components of the
macroscopic fields are zero when the argument is outside the
first Brillouin zone. This definition of macroscopic
fields is common in the physical literature 
\cite{Adler,Wiser,Scaife,Ehrenreich,Dutra,Langreth}.
Notice also that $\langle {\sf{f}}({\bf{R}},{\bf{\rho}})\rangle$
and $\langle \tilde{\sf{f}}({\bf{q}},{\bf{\rho}})\rangle$
are Fourier transforms of one another. Therefore, the macroscopic
fields ${\bf{E}}({\bf{R}})$ and ${\bf{E}}({\bf{q}})$
obtained by averaging the corresponding microscopic fields
remain Fourier transforms of one another.

By direct substitution, it can be shown that the
two-scale transforms of the gradient of $f$ are
$(\nabla_{\bf{R}} +\nabla_\rho) {\sf{f}} ({\bf{R}},{\bf{\rho}})$
and
$(i{\bf{q}}+\nabla_\rho) \tilde{\sf{f}} ({\bf{q}},{\bf{\rho}})$.

\section{The microscopic Maxwell equations}
To simplify the presentation, we consider a non magnetic sample and
we neglect the effect of spin (see Ref.\cite{Keller} for the
full theory). Moreover, the electromagnetic charges, currents
and fields
have a time dependence $\exp(-i\omega t)$ which
will be implicit for notational convenience.

The random phase approximation of quantum electrodynamics
corresponds to the following picture \cite{Ehrenreich}.
An external electromagnetic wave 
${\bf{E}}^{\mathrm{ext}}(\mathbf{r})$,
${\bf{B}}^{\mathrm{ext}}(\mathbf{r})$ polarizes
the dielectric crystal, creating a current density ${\bf{j}}({\bf{r}})$
and a charge density $n({\bf{r}})$. These current and charge
induce an electric field 
${\bf{e}}({\bf{r}})-{\bf{E}}^{\mathrm{ext}}(\mathbf{r})$
and a magnetic field 
${\bf{b}}({\bf{r}})-{\bf{B}}^{\mathrm{ext}}(\mathbf{r})$,
that induce additional current and charge, etc. When the medium 
and the field reach equilibrium, the vacuum (rationalized SI) 
Maxwell equations
describe the connection between the total fields and 
the induced charge and current densities:
\begin{eqnarray}
\nabla\cdot{\bf{e}}({\bf{r}}) &=& n({\bf{r}})/\epsilon_0
  \nonumber\\
\nabla\times{\bf{e}}({\bf{r}}) &=& i\omega{\bf{b}}({\bf{r}})
  \nonumber\\
\nabla\cdot{\bf{b}}({\bf{r}}) &=& 0
  \nonumber\\
\nabla\times{\bf{b}}({\bf{r}}) + i\omega\epsilon_0\mu_0{\bf{e}}({\bf{r}})
&=& \mu_0{\bf{j}}({\bf{r}}),
\end{eqnarray}
where charge conservation implies 
$\nabla\cdot{\bf{j}}({\bf{r}})=i\omega n({\bf{r}})$.

Linear response theory gives us the microscopic relation
\begin{equation}
{\bf{j}}({\bf{r}})=-i\omega\int d{\bf{r}}'
\chi({\bf{r}},{\bf{r}}')\cdot{\bf{e}}({\bf{r}}'),
\label{jchie}
\end{equation}
where the electric susceptibility $\chi_{ij}$ has the 
well-known expression
\cite{Keller,Agranovich,Negele,Ziman}
\begin{eqnarray}
\chi_{ij}({\bf{r}},{\bf{r}}')&=&
-\frac{e}{m\omega^2}
n_{0}(\mathbf{r})\delta_{ij}\delta(\mathbf{r}-\mathbf{r}')
  +\frac{e^2}{\omega^2}
  \sum_{n\not=0}
  \nonumber\\
&&\bigg[\frac{V^{0n}_i(\mathbf{r}) V^{n0}_j(\mathbf{r}')}
     {\hbar(\omega_{n0}-\omega)}+
\frac{V^{n0}_i(\mathbf{r}) V^{0n}_j(\mathbf{r}')}
     {\hbar(\omega_{n0}+\omega)}\bigg],
\label{chi}
\end{eqnarray}
where $n_{0}(\mathbf{r})$ is the charge density
in the ground state, $e$ is the 
(negative) electron charge and $m$ its
mass.  The velocity matrix elements are
\begin{eqnarray}
\mathbf{V}^{0n}({\mathbf{r}})&=&-\frac{i\hbar}{2m}
    [\Psi^*_0(\mathbf{r}) \nabla_r \Psi_n(\mathbf{r})
     - \Psi_n(\mathbf{r}) \nabla_r \Psi^*_0(\mathbf{r})],
\label{velocity}
\end{eqnarray}
where $\Psi_0(\mathbf{r})$ and $\Psi_n(\mathbf{r})$ are
eigenstates of the crystal. In the one-electron case, the
sum is carried out over the occupied ($0$) and 
unoccupied ($n$) states.

In expression (\ref{chi}), the first term, called 
the diamagnetic
term, depends only on the electronic density, and the
second (paramagnetic) term is usually much larger than
the first in the optical range \cite{Keller}.

Formula (\ref{chi}) for the susceptibility $\chi$ 
holds at zero temperature. The corresponding formula
for finite temperature involves a weighted sum over 
initial states \cite{Keller}. All the results of the
present paper can be straightforwardly adapted to
the finite temperature case.

From the microscopic current ${\bf{j}}({\bf{r}})$ it is 
customary in dielectric theory to define the polarization
${\bf{p}}({\bf{r}})=(i/\omega){\bf{j}}({\bf{r}})$ and
the displacement field
${\bf{d}}({\bf{r}})=\epsilon_0{\bf{e}}({\bf{r}})+
{\bf{p}}({\bf{r}})$. In terms of those fields, the
microscopic Maxwell equations become
\begin{eqnarray}
\nabla\times{\bf{e}}({\bf{r}}) &=& i\omega{\bf{b}}({\bf{r}})
  \nonumber\\
\nabla\cdot{\bf{b}}({\bf{r}}) &=& 0
  \nonumber\\
\nabla\cdot{\bf{d}}({\bf{r}}) &=& 0
  \nonumber\\
\nabla\times{\bf{b}}({\bf{r}}) &=& -i\omega\mu_0{\bf{d}}({\bf{r}})
\label{rotb},
\end{eqnarray}
where the microscopic constitutive relation is now
\begin{eqnarray*}
d_i({\bf{r}})=\epsilon_0 e_i({\bf{r}})+\sum_j\int d{\bf{r}}'
\chi_{ij}({\bf{r}},{\bf{r}}')e_j({\bf{r}}').
\end{eqnarray*}

The microscopic fields ${\bf{e}}({\bf{r}})$,
${\bf{b}}({\bf{r}})$ and ${\bf{p}}({\bf{r}})$ are,
as the susceptibility $\chi_{ij}({\bf{r}},{\bf{r}}')$, 
rapidly oscillating
functions of ${\bf{r}}$. The macroscopic fields
${\bf{E}}({\bf{R}})$, ${\bf{B}}({\bf{R}})$
and ${\bf{P}}({\bf{R}})$ are averages of the corresponding
quantities over a unit cell. The constitutive relations problem
is how to go from the relation between  ${\bf{p}}({\bf{r}})$
and  ${\bf{e}}({\bf{r}})$ to a relation between
${\bf{P}}({\bf{R}})$ and  ${\bf{E}}({\bf{R}})$.
Homogenization theory is a very convenient way to 
solve this problem.

\section{Homogenization of the microscopic Maxwell equations}
The derivation of the macroscopic Maxwell equations
from the microscopic ones is standard
\cite{Kranendonk,Scaife,Jackson}. Hence, the purpose of this
section is just to show that
homogenization theory gives the known results and
to derive equations that will be used in the next
sections.
The homogenization of the Maxwell equation has been
thoroughly studied by applied mathematicians
\cite{Bensoussan,Sanchez,Jikov,Sanchez3,Sanchez2,Amirat,Artola,Friedman,Markowich,Oster,Wellander}
and physicists \cite{Bossavit2,ElFeddi}. 
However, the microscopic constitutive relations
that they used were always local:
$\chi_{ij}({\bf{r}},{\bf{r}}')=
\chi_{ij}({\bf{r}})\delta({\bf{r}}-{\bf{r}}')$.
We do not need the full apparatus of 
homogenization theory
of non-local equations \cite{Lions} and we 
homogenize the Maxwell equations
by a simple adaptation of the method used for
local dielectric functions.

In this section, we follow the very clear homogenization
procedure of Sanchez-Palencia \cite{Sanchez}, except that
we work with ${\tilde{\sf{f}}}({\bf{q}},{\bf{\rho}})$
instead of ${\sf{f}}({\bf{R}},{\bf{\rho}})$.

Using the fact that the two-scale transform of
$\nabla f({\bf{r}})$ is 
$(i{\bf{q}}+\nabla_\rho) \tilde{\sf{f}} ({\bf{q}},{\bf{\rho}})$,
we can
make the two-scale transform
of the Maxwell equations (\ref{rotb}) to obtain:
\begin{eqnarray}
i{\bf{q}}\times{\bf{e}}({\bf{q}},{\bf{\rho}}) +
\nabla_\rho\times{\bf{e}}({\bf{q}},{\bf{\rho}}) &=& 
i\omega{\bf{b}}({\bf{q}},{\bf{\rho}})
  \nonumber\\
i{\bf{q}}\cdot{\bf{b}}({\bf{q}},{\bf{\rho}})+
\nabla_\rho\cdot{\bf{b}}({\bf{q}},{\bf{\rho}}) &=& 0
  \nonumber\\
i{\bf{q}}\cdot{\bf{d}}({\bf{q}},{\bf{\rho}})+
\nabla_\rho\cdot{\bf{d}}({\bf{q}},{\bf{\rho}}) &=& 0
  \nonumber\\
i{\bf{q}}\times{\bf{b}}({\bf{q}},{\bf{\rho}}) +
\nabla_\rho\times{\bf{b}}({\bf{q}},{\bf{\rho}}) &=& 
 -i\omega\mu_0{\bf{d}}({\bf{q}},{\bf{\rho}}).
\label{twoscaleMaxwell}
\end{eqnarray}
For notational convenience, we do not write 
the fields ${\bf{e}}({\bf{q}},{\bf{\rho}})$, etc,
with a tilde.

Because they vary slowly, the eventual ``external''
currents and charges would appear as
$\mathbf{j}(\mathbf{q})$ and $n(\mathbf{q})$,
with no dependence on $\rho$.

Consider a scattering problem where a plane
electromagnetic wave is shined on the dielectric.
In equations (\ref{twoscaleMaxwell}), the order of magnitude of 
${\bf{q}}$ will be $2\pi/\lambda$, where $\lambda$ is the
wavelength of the incident wave.
The order of magnitude of the unit cell dimensions 
is $l=|C|^{1/3}$.
Let $a=l/\lambda $, the order of magnitude of
$\nabla_\rho\cdot{\bf{e}}({\bf{q}},{\bf{\rho}})$
is $1/a$ times the order of magnitude of
${\bf{q}}\cdot{\bf{e}}({\bf{q}},{\bf{\rho}})$.
Now we expand all fields as sums of the type

\begin{eqnarray}
{\bf{e}}({\bf{q}},{\bf{\rho}})=
{\bf{e}}^{(0)}({\bf{q}},{\bf{\rho}})+
a{\bf{e}}^{(1)}({\bf{q}},{\bf{\rho}})+
a^2{\bf{e}}^{(2)}({\bf{q}},{\bf{\rho}})+\cdots
\label{asumtype}
\end{eqnarray}
where all terms of the expansion are periodic in ${\bf{\rho}}$.
Since $a$ is small, we keep only the first term of the
expansion to define the macroscopic fields as
${\bf{E}}({\bf{q}})=\langle {\bf{e}}^{(0)}({\bf{q}},{\bf{\rho}})\rangle$,
etc.
The validity \cite{Developpement} and asymptotic
convergence of this expansion is the main technical difficulty
that was solved by the mathematicians who homogenized the
Maxwell equations (see, for instance, Ref.\cite{Friedman}
where the general term of expansion (\ref{asumtype}) is given).
To be complete, we need to know that the order of
magnitude of $\omega$ is $2\pi c/\lambda$, the
order of manitude of $\mathbf{b}$ is $\mathbf{e}/c$ and
the order of magnitude of $\mathbf{d}$ is $\epsilon_0\mathbf{e}$.

Introducing the expansions (\ref{asumtype})
into Eq.(\ref{twoscaleMaxwell}) and gathering
all terms of same power of $a$ we obtain four equations
for the $a^{-1}$ term:
\begin{eqnarray}
\nabla_\rho\times{\bf{e}}^{(0)}({\bf{q}},{\bf{\rho}}) &=& 0 
\nonumber\\
\nabla_\rho\cdot{\bf{b}}^{(0)}({\bf{q}},{\bf{\rho}}) &=& 0
  \nonumber\\
\nabla_\rho\cdot{\bf{d}}^{(0)}({\bf{q}},{\bf{\rho}}) &=& 0
 \label{drhoq} \\
\nabla_\rho\times{\bf{b}}^{(0)}({\bf{q}},{\bf{\rho}}) &=& 0,
\nonumber
\end{eqnarray}
and four equations for the $a^{0}$ term:
\begin{eqnarray}
i{\bf{q}}\times{\bf{e}}^{(0)}({\bf{q}},{\bf{\rho}}) +
a\nabla_\rho\times{\bf{e}}^{(1)}({\bf{q}},{\bf{\rho}}) &=& 
i\omega{\bf{b}}^{(0)}({\bf{q}},{\bf{\rho}})
  \nonumber\\
i{\bf{q}}\cdot{\bf{b}}^{(0)}({\bf{q}},{\bf{\rho}})+
a\nabla_\rho\cdot{\bf{b}}^{(1)}({\bf{q}},{\bf{\rho}}) &=& 0
  \label{divb1}\\
i{\bf{q}}\cdot{\bf{d}}^{(0)}({\bf{q}},{\bf{\rho}})+
a\nabla_\rho\cdot{\bf{d}}^{(1)}({\bf{q}},{\bf{\rho}}) &=& 0
  \nonumber\\
i{\bf{q}}\times{\bf{b}}^{(0)}({\bf{q}},{\bf{\rho}}) +
a\nabla_\rho\times{\bf{b}}^{(1)}({\bf{q}},{\bf{\rho}}) &=& 
 -i\omega\mu_0{\bf{d}}^{(0)}({\bf{q}},{\bf{\rho}})\nonumber,
\end{eqnarray}
(recall that $a\nabla_\rho$ has the same order of
magnitude as $\mathbf{q}$).
These two sets of equations are sufficient to determine the
macroscopic Maxwell equations. Further terms would be
required if the wavelength were not very large
with respect to the unit cell dimensions.

To obtain the macroscopic Maxwell equations, we must eliminate
the terms of order one, e.g. ${\bf{b}}^{(1)}({\bf{q}},{\bf{\rho}})$.
This is achieved by transforming the cell average of the
divergence into an integral over the surface $\partial C$
of the unit cell:
$\langle\nabla_\rho\cdot{\bf{b}}^{(1)}({\bf{q}},{\bf{\rho}})
\rangle=1/{|C|}
\int_{\partial C} \hat{n}\cdot
{\bf{b}}^{(1)}({\bf{q}},{\bf{\rho}})dS$. On opposite
sides of cell $C$, the outgoing normal $\hat{n}$ is reversed
while ${\bf{b}}^{(1)}({\bf{q}},{\bf{\rho}})$ is
equal (by periodicity). Therefore, the surface integral is zero
and 
$\langle\nabla_\rho\cdot{\bf{b}}^{(1)}({\bf{q}},{\bf{\rho}})
\rangle=0$.
A similar reasoning leads to 
$\langle\nabla_\rho\times{\bf{b}}^{(1)}({\bf{q}},{\bf{\rho}})
\rangle=0$.

Thus, averaging Eq.({\ref{divb1}) gives
$0=i{\bf{q}}\cdot\langle{\bf{b}}^{(0)}({\bf{q}},{\bf{\rho}})\rangle+
a\langle \nabla_\rho\cdot{\bf{b}}^{(1)}({\bf{q}},{\bf{\rho}})\rangle
=i{\bf{q}}\cdot{\bf{B}}({\bf{q}})$, because the
macroscopic field ${\bf{B}}({\bf{q}})$ was defined 
as the average of ${\bf{b}}^{(0)}({\bf{q}},{\bf{\rho}})$.
Carrying out similar
calculations for the other microscopic Maxwell equations we obtain the
macroscopic Maxwell equations in momentum space:
\begin{eqnarray}
i{\bf{q}}\times\bf{E}({\bf{q}}) &=& i\omega\bf{B}({\bf{q}})
  \nonumber\\
i{\bf{q}} \cdot\bf{B}({\bf{q}}) &=& 0
  \nonumber\\
i{\bf{q}}\cdot\bf{D}({\bf{q}}) &=& 0
  \nonumber\\
i{\bf{q}}\times\bf{B}({\bf{q}}) &=& -i\omega\mu_0\bf{D}({\bf{q}}).
\end{eqnarray}
Fourier transforming back to real space we obtain the
usual macroscopic Maxwell equations:
\begin{eqnarray}
\nabla\times\bf{E}({\bf{R}}) &=& i\omega\bf{B}({\bf{R}})
  \nonumber\\
\nabla\cdot\bf{B}({\bf{R}}) &=& 0
  \nonumber\\
\nabla\cdot\bf{D}({\bf{R}}) &=& 0
  \nonumber\\
\nabla\times\bf{B}({\bf{R}}) &=& -i\omega\mu_0\bf{D}({\bf{R}})
\nonumber.
\end{eqnarray}

Notice that this homogenization procedure can be applied 
directly
to Eqs.(\ref{twoscaleMaxwell}), but the present formulation
is required for consistency with the derivation of the
constitutive relations.

Homogenization now proceeds classically. We start by 
homogenizing the
magnetic field ${\bf{b}}^{(0)}({\bf{q}},{\bf{\rho}})$.
From the equation 
$\nabla_\rho\times{\bf{b}}^{(0)}({\bf{q}},{\bf{\rho}})=0$
we deduce that there is a periodic potential
$\phi({\bf{q}},{\bf{\rho}})$ and a function of ${\bf{q}}$,
denoted by ${\bf{F}}({\bf{q}})$, such that
${\bf{b}}^{(0)}({\bf{q}},{\bf{\rho}})={\bf{F}}({\bf{q}})
-\nabla_\rho\phi({\bf{q}},{\bf{\rho}})$. Taking the average
of both sides, and considering the periodicity of $\phi$
we obtain 
$\langle {\bf{b}}^{(0)}({\bf{q}},{\bf{\rho}})\rangle=
{\bf{F}}({\bf{q}})$ \cite{intder}
and, since the left-hand side is defined
as the macroscopic magnetic field ${\bf{B}}({\bf{q}})$,
we get 
${\bf{b}}^{(0)}({\bf{q}},{\bf{\rho}})={\bf{B}}({\bf{q}})
-\nabla_\rho\phi({\bf{q}},{\bf{\rho}})$.
Introducing this equality into 
$\nabla_\rho\cdot{\bf{b}}^{(0)}({\bf{q}},{\bf{\rho}})=0$
we obtain the equation 
$\Delta_\rho\phi({\bf{q}},{\bf{\rho}})=0$.
The only periodic solution of the Laplace equation is a
constant \cite{Bensoussan}, so
$\phi({\bf{q}},{\bf{\rho}})$ is a constant and
${\bf{b}}^{(0)}({\bf{q}},{\bf{\rho}})={\bf{B}}({\bf{q}})$.
In other words, the magnetic field does not need to be averaged,
the zero-th order term of 
${\bf{b}}({\bf{q}},{\bf{\rho}})$ is smooth and does not depend
on $\rho$. The periodic modulation is only reached at the next and
smaller term $a{\bf{b}}^{(1)}({\bf{q}},{\bf{\rho}})$.

Such an automatic averaging is not possible for 
the electric field. Starting from
$\nabla_\rho\times{\bf{e}}^{(0)}({\bf{q}},{\bf{\rho}})=0$,
the only conclusion that we can reach at this level
is that there is a periodic potential 
$\phi({\bf{q}},{\bf{\rho}})$ such that
${\bf{e}}^{(0)}({\bf{q}},{\bf{\rho}})={\bf{E}}({\bf{q}})-
\nabla_\rho\phi({\bf{q}},{\bf{\rho}})$. The determination of
$\phi$ will be the purpose of the next sections, but
this result has already an interesting meaning. The
zero-th term of the expansion of 
${\bf{e}}({\bf{q}},{\bf{\rho}})$ is not smooth.
In mathematical terms,
the limit of ${\bf{e}}({\bf{q}},{\bf{\rho}})$ as
$a\rightarrow 0$ is still an oscillating function,
contrary to our example $f(x/a)=\sin(x/a)+b$.
This concept of ``oscillating limit'' was introduced
by the Cameroonese mathematician G. Nguetseng in 1989
\cite{Nguetseng}
and has deeply simplified homogenization theory
\cite{Allaire}.

\section{The constitutive relation}
Still, to give a complete description of the
macroscopic electromagnetic properties of matter,
we have to establish a correspondence between
$\bf{D}({\bf{q}})$ and $\bf{E}({\bf{q}})$.

To do this, we use the periodicity of 
$\chi$: for every
lattice vector ${\bf{R}}$,
$\chi_{ij}({\bf{r}}+{\bf{R}},{\bf{r}}'+{\bf{R}})=
\chi_{ij}({\bf{r}},{\bf{r}}')$ \cite{Agranovich}.
Notice that the same lattice vector must be added
to both arguments of $\chi$.

We start from the relation between the polarization
and the electric field
\begin{eqnarray*}
{\bf{p}}({\bf{r}}) &=& \int d{\bf{r}}'
 \chi({\bf{r}},{\bf{r}}')\cdot{\bf{e}}({\bf{r}}').
\end{eqnarray*}
Then, we apply the two-scale transformation
Eq.(\ref{deraqrho}) to find
\begin{eqnarray*}
{\bf{p}}({\bf{q}},{\bf{\rho}}) &=& 
\frac{|C|}{(2\pi)^3} \sum_s 
\exp[-i{\bf{q}}\cdot({\bf{\rho}}+{\bf{R}}_s)]\\&&
\int d{\bf{r}}'
 \chi({\bf{\rho}}+{\bf{R}}_s,{\bf{r}}')\cdot{\bf{e}}({\bf{r}}').
\end{eqnarray*}
Now the integral over all space is split into integrals over
translated unit cells:
\begin{eqnarray*}
{\bf{p}}({\bf{q}},{\bf{\rho}}) &=& 
\frac{|C|}{(2\pi)^3} \sum_{s s'}
\exp[-i{\bf{q}}\cdot({\bf{\rho}}+{\bf{R}}_s)]\\&&
\int_C d{\bf{\rho}}'
 \chi({\bf{\rho}}+{\bf{R}}_s,{\bf{\rho}}'+{\bf{R}}_{s'})
 \cdot{\bf{e}}({\bf{\rho}}'+{\bf{R}}_{s'}).
\end{eqnarray*}
If we introduce the two-scale expression (\ref{fder}) for the
electric field
${\bf{e}}({\bf{\rho}})=\int_{BZ} d{\bf{q}}
  \exp(i{\bf{q}}\cdot{\bf{\rho}}) {{\bf{e}}}({\bf{q}},{\bf{\rho}})$
and use the periodicity of 
${{\bf{e}}}({\bf{q}},{\bf{\rho}})$ in ${\bf{\rho}}$ we get
\begin{eqnarray*}
{\bf{p}}({\bf{q}},{\bf{\rho}}) &=& 
\frac{|C|}{(2\pi)^3} \sum_{s s'}
\exp[-i{\bf{q}}\cdot({\bf{\rho}}+{\bf{R}}_s)]
\int_{BZ} d{\bf{q}}' \int_C d{\bf{\rho}}' \\&&\hspace*{-5mm}
\exp[i{\bf{q}}'\cdot({\bf{\rho}}'+{\bf{R}}_{s'})]
 \chi({\bf{\rho}}+{\bf{R}}_s,{\bf{\rho}}'+{\bf{R}}_{s'})
 \cdot{{\bf{e}}}({\bf{q}}',{\bf{\rho}}').
\end{eqnarray*}
If we replace ${\bf{R}}_s$ by
${\bf{R}}_{t}+{\bf{R}}_{s'}$ and use the periodicity of $\chi$
we obtain
\begin{eqnarray*}
{\bf{p}}({\bf{q}},{\bf{\rho}}) &=& 
\frac{|C|}{(2\pi)^3} \sum_{t}
\exp[-i{\bf{q}}\cdot({\bf{\rho}}+{\bf{R}}_t)]
\int_{BZ} d{\bf{q}}'\\&&\int_C d{\bf{\rho}}'
\exp(i{\bf{q}}'\cdot{\bf{\rho}}')
\sum_{s'}
\exp[i({\bf{q}}'-{\bf{q}})\cdot{\bf{R}}_{s'}]\\&&
 \chi({\bf{\rho}}+{\bf{R}}_t,{\bf{\rho}}')
 \cdot{{\bf{e}}}({\bf{q}}',{\bf{\rho}}').
\end{eqnarray*}
The sum over $s'$ is carried out with
Eq.(\ref{somsurs}) and, since ${\bf{q}}$
and ${\bf{q}}'$ belong to the first Brillouin
zone, only the term ${\bf{K}}=0$ contributes.
We reach finally
\begin{eqnarray*}
{\bf{p}}({\bf{q}},{\bf{\rho}}) &=& 
\sum_{s}\int_C d{\bf{\rho}}'
\exp[-i{\bf{q}}\cdot({\bf{\rho}}-{\bf{\rho}}'+{\bf{R}}_s)]\\&&
\chi({\bf{\rho}}+{\bf{R}}_s,{\bf{\rho}}')
 \cdot{{\bf{e}}}({\bf{q}},{\bf{\rho}}')\\
&=& \langle \tilde{\chi}({\bf{\rho}},{\bf{\rho}}'; {\bf{q}})
\cdot{{\bf{e}}}({\bf{q}},{\bf{\rho}}')\rangle_{\rho'},
\end{eqnarray*}
where the index ${\rho'}$ designates cell average
over variable ${\rho'}$, and where
we have defined the two-scale transform of $\chi$
as
\begin{eqnarray}
\tilde{\chi}({\bf{\rho}},{\bf{\rho}}'; {\bf{q}}) &=& 
|C|\sum_{s}
\exp[-i{\bf{q}}\cdot({\bf{\rho}}-{\bf{\rho}}'+{\bf{R}}_s)]
\chi({\bf{\rho}}+{\bf{R}}_s,{\bf{\rho}}').\nonumber\\
&&\label{twoscalechi}
\end{eqnarray}

It can be checked that 
$\tilde{\chi}({\bf{\rho}},{\bf{\rho}}'; {\bf{q}})$
has the lattice periodicity for each
variable ${\bf{\rho}}$ and ${\bf{\rho}}'$
independently. A related definition was used by
Ehrenreich \cite{Ehrenreich}.

All quantities are now periodic and diagonal in ${\bf{q}}$:
they are in a suitable form for homogenization. From the
relation between polarization and electric field, we deduce
the relation between displacement and electric field which
will be our starting point:
\begin{eqnarray}
d_i({\bf{q}},{\bf{\rho}})=
\epsilon_0 e_i({\bf{q}},{\bf{\rho}})
+\sum_j\langle
\tilde\chi_{ij}({\bf{\rho}},{\bf{\rho}}';{\bf{q}})
e_j({\bf{q}},{\bf{\rho}}')\rangle_{\rho'}.\label{macro1}
\end{eqnarray}

The macroscopic constitutive relation is obtained by
restricting all fields to the first term of 
expansion (\ref{asumtype}). If we do this in
Eq.(\ref{macro1}) and write
the electric field
as the sum
${\bf{e}}^{(0)}({\bf{q}},{\bf{\rho}})=
{\bf{E}}({\bf{q}})-\nabla_\rho\phi({\bf{q}},{\bf{\rho}})$
we obtain
\begin{eqnarray}
d^{(0)}_i({\bf{q}},{\bf{\rho}})&=&
\epsilon_0 E_i({\bf{q}})+
\sum_j\langle
\tilde\chi_{ij}({\bf{\rho}},{\bf{\rho}}';{\bf{q}})\rangle_{\rho'}
E_j({\bf{q}})-
\nonumber\\&&\hspace*{-12mm}
\epsilon_0 \partial_{\rho_i}\phi({\bf{q}},{\bf{\rho}})
-\sum_j\langle
\tilde\chi_{ij}({\bf{\rho}},{\bf{\rho}}';{\bf{q}})
\partial_{\rho'_j}\phi({\bf{q}},{\bf{\rho}}')\rangle_{\rho'}.
\label{macro2}
\end{eqnarray}
If Eq.(\ref{macro2}) is averaged over $\rho$, 
the fact that
$\langle \partial_{\rho_i}\phi({\bf{q}},{\bf{\rho}})\rangle=0$
\cite{intder} leads to
the first step of the macroscopic constitutive
relation:
\begin{eqnarray}
D_i({\bf{q}})&=&
\epsilon_0 E_i({\bf{q}})
+\sum_j\langle
\tilde\chi_{ij}({\bf{\rho}},{\bf{\rho}}';{\bf{q}})\rangle_{\rho\rho'}
E_j({\bf{q}})
\nonumber\\&&
-\sum_j\langle
\tilde\chi_{ij}({\bf{\rho}},{\bf{\rho}}';{\bf{q}})
\partial_{\rho'_j}\phi({\bf{q}},{\bf{\rho}}')\rangle_{\rho\rho'}.
\label{macro3}
\end{eqnarray}

Following Wiser \cite{Wiser},
a connexion with the classical approach is possible 
through the introduction of a macroscopic 
local field $\mathbf{E}_{\mathrm{loc}}(\mathbf{q})$ defined by
\begin{eqnarray*}
\langle\tilde\chi({\bf{\rho}},{\bf{\rho}}';{\bf{q}})
\rangle_{\rho\rho'}\cdot
\mathbf{E}_{\mathrm{loc}}(\mathbf{q})=
\langle
\tilde\chi({\bf{\rho}},{\bf{\rho}}';{\bf{q}})\cdot
\mathbf{e}({\bf{q}},{\bf{\rho}}')\rangle_{\rho\rho'},
\end{eqnarray*}
so that
$\mathbf{D}(\mathbf{q})=\epsilon_0\mathbf{E}(\mathbf{q})
+\langle\tilde\chi({\bf{\rho}},{\bf{\rho}}';{\bf{q}})
\rangle_{\rho\rho'}\cdot
\mathbf{E}_{\mathrm{loc}}(\mathbf{q})$.

The next step is the determination of
$\phi({\bf{q}},{\bf{\rho}}')$.
We take the microscopic equation 
$\nabla_\rho\cdot{\bf{d}}^{(0)} ({\bf{q}},{\bf{\rho}})=0$
derived in the previous section (Eq.(\ref{drhoq})),
and we apply it
to Eq.(\ref{macro2}). We obtain an equation for 
$\phi({\bf{q}},{\bf{\rho}})$:
\begin{eqnarray}
&&\epsilon_0\Delta_\rho\phi({\bf{q}},{\bf{\rho}})+
\sum_{ij}\langle\partial_{\rho_i}
\tilde\chi_{ij}({\bf{\rho}},{\bf{\rho}}';{\bf{q}})
\partial_{\rho'_j}\phi({\bf{q}},{\bf{\rho}}')\rangle_{\rho'}=
\nonumber\\&& \hspace{6mm}
\sum_{ij}\langle\partial_{\rho_i}
\tilde\chi_{ij}({\bf{\rho}},{\bf{\rho}}';{\bf{q}})\rangle_{\rho'}
E_j({\bf{q}}). \label{cell}
\end{eqnarray}
This so-called cell equation determines a unique periodic solution
$\phi({\bf{q}},{\bf{\rho}})$ with zero average over a period.

We follow first the standard argument of
homogenization theory.
Let the three functions $A_k({\bf{q}},{\bf{\rho}})$ 
($k=x,y,z$) be the solutions of the cell
equation for an electric field ${\bf{E}}$ equal
to a unit vector in the direction $k$. Then, for
a general electric field ${\bf{E}} ({\bf{q}})$, the
potential is
$\phi({\bf{q}},{\bf{\rho}})={\bf{A}}({\bf{q}},{\bf{\rho}})\cdot
{\bf{E}} ({\bf{q}})$.

If we introduce this expression for 
$\phi({\bf{q}},{\bf{\rho}})$ 
into Eq.(\ref{macro3}),
we obtain the macroscopic constitutive relation
$D_i({\bf{q}})=\sum_j \epsilon_{ij}({\bf{q}})
E_j({\bf{q}})$ where the macroscopic dielectric constant
$\epsilon_{ij}({\bf{q}})$ is given by
\begin{eqnarray*}
\epsilon_{ij}({\bf{q}})&=&\epsilon_0\delta_{ij}
+\sum_k\langle
\tilde\chi_{ik}({\bf{\rho}},{\bf{\rho}}';{\bf{q}})
[\delta_{kj}-\partial_{\rho'_k} A_j({\bf{q}},{\bf{\rho}}')]
\rangle_{\rho \rho'}.
\end{eqnarray*}

This equation shows that the macroscopic 
dielectric constant is the average of the
microscopic dielectric function plus a correction term.
In the case when the dielectric function is a sum of
constant factors localized at 
all sites of a cubic lattice, it is standard exercice
of homogenization theory to show that the macroscopic
constitutive relation
becomes the Clausius-Mossotti equation (see Ref.\cite{Jikov},
p.45). Therefore, the correction term can be quite large
and should not be neglected.

Homogenization usually stops here, and the integro-differential
cell equation (\ref{cell})
can only be solved numerically, which seems to be a 
reasonable task. For instance,
the FLAPW approach could be used, where space is cut
into non-overlapping 
spheres plus an interstitial region. $\phi({\bf{q}},{\bf{\rho}})$
is then expanded over spherical harmonics (with suitable radial
functions) in the spheres
and over plane waves $\exp(i\mathbf{K}\cdot\rho)$ in the 
interstitial region. This reduces the cell equation to 
a matrix equation.

However, the particular 
structure of $\chi_{ij}({\bf{r}},{\bf{r}}')$ can be used
to give an explicit solution for $\phi({\bf{q}},{\bf{\rho}})$.
This is the purpose of the next section.

\section{The cell problem}
The cell equation (\ref{cell}) can be interpreted as follows.
If a constant electric field ${\bf{E}}({\bf{q}})$
is applied to the dielectric, linear response theory 
tells us that it induces a periodic current
given by Eq. (\ref{jchie})
$$
{\bf{j}}({\bf{q}},{\bf{\rho}})=
-i\omega\langle \tilde\chi({\bf{\rho}},{\bf{\rho}}';{\bf{q}})
\rangle_{\rho'}
\cdot {\bf{E}}({\bf{q}})
$$
and the corresponding periodic charge is
\begin{eqnarray*}
n({\bf{q}},{\bf{\rho}})&=&-(i/\omega)\nabla_\rho\cdot
{\bf{j}}({\bf{q}},{\bf{\rho}})\\&=&
- \nabla_\rho\cdot\langle \tilde\chi({\bf{\rho}},{\bf{\rho}}';{\bf{q}})
\rangle_{\rho'}
\cdot {\bf{E}}({\bf{q}}).
\end{eqnarray*}
The last term is minus the right-hand side of
Eq.(\ref{cell}).
This periodic charge induces an additional electric
field, that creates an additional polarization, and 
$\phi({\bf{q}},{\bf{\rho}})$ is the periodic potential
(with zero average) which represents the local electric field
reached at equilibrium under the influence of the external
field ${\bf{E}}({\bf{q}})$. In other words, the cell
equation (\ref{cell}) is the electrostatic equation
for the potential created in a unit cell of the dielectric
by an external field ${\bf{E}}({\bf{q}})$.

With this picture in mind, we can solve the cell equation
iteratively. We need the periodic electrostatic Green
function $G^{\#}({\bf{\rho}})$ which is a solution of
$\epsilon_0\Delta G^{\#}({\bf{\rho}})=-\delta({\bf{\rho}})$ in a unit cell.
Some properties of $G^{\#}({\bf{\rho}})$ are
discussed in Ref.\cite{Jikov}, p.121, 
\begin{eqnarray}
G^{\#}({\bf{\rho}})&=&\sum_{{\bf{K}}}\frac{\exp(i{\bf{K}}\cdot\rho)}
  {\epsilon_0 |C|(|{\bf{K}}|^2-i\eta)}\label{gdieseK}\\
&=&\sum_s\frac{1}{4\pi\epsilon_0|\rho+{\bf{R}}_s|}\nonumber,
\end{eqnarray}
where $\eta$ is an infinitesimal positive real.

If the Green function is applied to both
sides of Eq.(\ref{cell}) we obtain
\begin{eqnarray*}
\phi({\bf{q}},{\bf{\rho}})&=&-\frac{1}{|C|}\int d\tau d\rho'
G^{\#}(\rho-\tau)\nabla_{\tau}\cdot
\chi({\bf{\tau}},{\bf{\rho}}';{\bf{q}})\cdot
\mathbf{E}({\bf{q}})\\&&\hspace*{-5mm}+
\frac{1}{|C|}\int d\tau d\rho'
G^{\#}(\rho-\tau)\nabla_{\tau}\cdot
\chi({\bf{\tau}},{\bf{\rho}}';{\bf{q}})\cdot
\nabla_{\rho'}\phi({\bf{q}},{\bf{\rho}}').
\end{eqnarray*}

An iterative solution of this equation 
can be written, in simplified notation
\begin{eqnarray*}
\phi=-\frac{1}{|C|} G^{\#}\nabla\cdot\tilde\chi\cdot{\bf{E}}-
 \frac{1}{|C|^2} G^{\#}\nabla\cdot\tilde\chi\cdot\nabla 
G^{\#}\nabla\cdot\tilde\chi\cdot {\bf{E}}+\cdots
\end{eqnarray*}
If we introduce our iterative solution in Eq.(\ref{macro3})
we obtain
\begin{eqnarray}
{\bf{D}}= \epsilon_0{\bf{E}}+\langle \tilde\chi+ \frac{1}{|C|}
\tilde\chi\cdot\nabla G^{\#}\nabla\cdot\tilde\chi+\cdots 
\rangle \cdot{\bf{E}}.
\label{macro4}
\end{eqnarray}
To be more explicit, we introduce the macroscopic
susceptibility $\bar\chi_{ij}(\mathbf{q})$ and
write Eq.(\ref{macro4}) as
${\bf{D}}(\mathbf{q})= \epsilon_0{\bf{E}}(\mathbf{q})+
\bar\chi(\mathbf{q})\cdot{\bf{E}}(\mathbf{q})$, where
\begin{eqnarray}
\bar\chi_{ij}(\mathbf{q})
&=&\frac{1}{|C|^2}\int_{C\times C}d\mathbf{\rho}d\mathbf{\rho'}
\tilde\chi_{ij}(\mathbf{\rho},\mathbf{\rho}';\mathbf{q})
\nonumber\\&&+
\frac{1}{|C|^3}\sum_{mn}
\int_{C^4}d\mathbf{\rho}d\mathbf{\rho}'
    d\mathbf{\tau}d\mathbf{\tau}'
\tilde\chi_{im}(\mathbf{\rho},\mathbf{\tau};\mathbf{q})\nonumber\\&&
\partial_{\tau_m}G^{\#}(\tau-\tau')\partial_{\tau'_n}
\tilde\chi_{nj}(\mathbf{\tau}',\mathbf{\rho}';\mathbf{q})+\cdots
\label{macro4bis}
\end{eqnarray}
or
\begin{eqnarray*}
\bar\chi(\mathbf{q})
&=&\langle\tilde\chi(\mathbf{q})\cdot
[1-\frac{1}{|C|}\nabla G^{\#}\nabla\cdot
\tilde\chi(\mathbf{q})]^{-1}
\rangle_{\rho\rho'}.
\end{eqnarray*}

To sum the right-hand side of Eq.(\ref{macro4}), we need
a separable form for $\tilde\chi$, akin to work of
Cho \cite{Cho} or the coupled-antenna
theory of Keller \cite{Keller}.
If we neglect the diamagnetic component of $\chi$
in Eq.(\ref{chi}), we can write it in the separable form
\begin{eqnarray}
\tilde\chi_{ij}({\bf{\rho}},{\bf{\rho}}';{\bf{q}})=
\sum_n f^n_i({\bf{\rho}},{\bf{q}}) g^n_j({\bf{\rho}}',{\bf{q}}).
\label{sepchi}
\end{eqnarray}
Introducing this representation in Eq.(\ref{macro4}),
everything decouples and we obtain
$D_i({\bf{q}})=\sum_j \epsilon_{ij}({\bf{q}})
E_j({\bf{q}})$ where the macroscopic dielectric constant
$\epsilon_{ij}({\bf{q}})$ is now given by
\begin{eqnarray}
\epsilon_{ij}({\bf{q}})&=&\epsilon_0\delta_{ij}
+\sum_{nn'}
\langle f^n_i({\bf{\rho}},{\bf{q}})\rangle
{[1-M({\bf{q}})]}^{-1}_{nn'} 
\langle g^{n'}_j({\bf{\rho}},{\bf{q}})\rangle
,\nonumber\\&&
\label{macro5}
\end{eqnarray}
and where the reaction field matrix 
(or screening matrix \cite{Hanke}) is defined as
\begin{eqnarray}
M({\bf{q}})_{nn'}&=&|C|\sum_{ij}
\langle g^n_i({\bf{\rho}},{\bf{q}})\partial_{\rho_i}
G^{\#}({\bf{\rho}}-{\bf{\rho}}') \partial_{\rho'_j}
f^{n'}_j ({\bf{\rho}}',{\bf{q}})\rangle_{\rho\rho'}.
\nonumber\\&&
\label{Meq}
\end{eqnarray}

Equations (\ref{macro5}) and (\ref{Meq}), giving an explicit 
expression for the macroscopic dielectric
constant, are the main result of the paper.
These equations can be useful when only a few states $n$
contribute to the susceptibility
$\chi(\rho,\rho';\mathbf{q})$

Integrating Eq.(\ref{Meq}) by parts and using 
the periodicity of all functions involved, 
we can derive an alternative form for
the reaction field matrix $M({\bf{q}})_{nn'}$,
which decreases the singularity of the derivative of 
$G^{\#}$:
\begin{eqnarray*}
M({\bf{q}})_{nn'}=-|C|
\langle 
G^{\#}({\bf{\rho}}-{\bf{\rho}}')
\nabla_\rho\cdot\mathbf{g}^n({\bf{\rho}},{\bf{q}})
\nabla_{\rho'}\cdot\mathbf{f}^{n'}({\bf{\rho}}',{\bf{q}})
\rangle_{\rho\rho'}.
\end{eqnarray*}

An alternative summation of the series (\ref{macro4})
can be obtained by separating $G^{\#}({\bf{\rho}}-{\bf{\rho}}')$
with Eq.(\ref{gdieseK}). A calculation similar to the 
foregoing one leads to
\begin{eqnarray*}
\epsilon_{ij}({\bf{q}})&=&\epsilon_0\delta_{ij}
+\langle\tilde\chi_{ij}(\rho,\rho';\mathbf{q})\rangle_{\rho\rho'}
\\&&\hspace*{-12mm}
+{\sum_{\mathbf{K},\mathbf{K}'}}'
 \frac{1}{\epsilon_0|\mathbf{K}||\mathbf{K}'|}
 \sum_{mn} 
 \langle \exp(i\mathbf{K}\cdot\rho')
 \partial_{\rho'_m}\tilde\chi_{im}(\rho,\rho';\mathbf{q})\rangle_{\rho\rho'}
\\&&
  (1-N)^{-1}_{\mathbf{K}\mathbf{K}'}
  \langle \exp(-i\mathbf{K}'\cdot\rho)
  \partial_{\rho_n}\tilde\chi_{nj}(\rho,\rho';\mathbf{q})\rangle_{\rho\rho'},
\end{eqnarray*}
where the reaction field matrix is now
\begin{eqnarray}
N_{\mathbf{K}\mathbf{K}'}&=&\frac{1}{\epsilon_0|\mathbf{K}||\mathbf{K}'|}
  \sum_{ij} \langle \exp(-i\mathbf{K}\cdot\rho)\\&&
 \hspace*{-6mm}
  [\partial_{\rho_i} \partial_{\rho'_j}
  \tilde\chi_{ij}(\rho,\rho';\mathbf{q})]
  \exp(i\mathbf{K}'\cdot\rho')\rangle_{\rho\rho'}.
 \label{NKK}
\end{eqnarray}

The notation $\sum'$ means that the sum is over all 
{\em non-zero} reciprocal lattice vectors.
To show that the terms $\mathbf{K}=0$ or $\mathbf{K}'=0$
do no contribute, we reintroduce the infinitesimal number
$-i\eta$ of Eq.(\ref{gdieseK}) and, for instance,
the $\mathbf{K}'=0$ term gives us
$(i/\eta)\langle\nabla_{\rho}\cdot
\tilde\chi(\rho,\rho';\mathbf{q})\rangle_{\rho\rho'}=0$,
because the average of a divergence is zero \cite{intder}.
Similarly, $N_{\mathbf{K}\mathbf{K}'}$ is
zero for $\mathbf{K}=0$ or $\mathbf{K}'=0$.
The last expression for $\epsilon(\mathbf{q})$
is computationnaly effective when 
$\chi(\rho,\rho';\mathbf{q})$
is smooth and only a few $\mathbf{K}$ contribute.

Notice that Eq.(\ref{NKK}) amounts to using the
Fourier transform of the susceptibility 
$\tilde\chi_{ij}(\rho,\rho';\mathbf{q})$.
We do not discuss this approach further, since it
has been used by many authors 
\cite{Adler,Wiser,Mochan2,Sinha,Mochan}.
It can be checked that Eq.(\ref{NKK}) is 
equivalent their results, in the limit
where $a\rightarrow 0$.

\section{Basic properties}
In this section, some basic consequences of the
macroscopic constitutive relations are derived.
Firstly, we show how the usual concept of
electric dipole transition is recovered, then
that the macroscopic constitutive relations
are causal and have the required symmetry properties.
Finally we prove that the average of the
microscopic energy density is given by
the macroscopic energy density and
we discuss the reaction field matrix.

\subsection{Electric dipole transitions}
To have a constitutive relation in the real space, 
we back Fourier transform the equation
$\mathbf{D}(\mathbf{q})=\epsilon(\mathbf{q})\cdot
\mathbf{E}(\mathbf{q})$ and we obtain
\begin{eqnarray}
D_i({\bf{r}})=\sum_j \int d{\bf{r}}'
\bar\epsilon_{ij}({\bf{r}}-{\bf{r}}')
E_j({\bf{r}}'), \label{homogen}
\end{eqnarray}
where 
\begin{eqnarray}
\bar\epsilon_{ij}({\bf{r}}-{\bf{r}}')=\frac{1}{(2\pi)^3}
\int d{\bf{q}}
\exp[i{\bf{q}}\cdot({\bf{r}}-{\bf{r}}')]
\epsilon_{ij}({\bf{q}}).\label{epsder} 
\end{eqnarray}

Equation (\ref{homogen}) is typical of a
homogeneous (but generally anisotropic) medium
\cite{Agranovich}. The fact that Eq.(\ref{homogen})
is non-local corresponds to spatial dispersion,
which has been much studied by the Russian school \cite{Agranovich},
and has received renewed interest recently
\cite{Morro,Koopmans,Graham,Gunning}. 
Besides, the $\mathbf{q}$-dependence of
$\epsilon_{ij}({\bf{q}})$ can be observed 
experimentally by inelastic electron, x-ray or
neutron scattering \cite{Tarrio,Tarrio2,Li}.
In Eq.(\ref{epsder}), it is not necessary to
restrict the integral to the first Brillouin
zone since, by definition, $\epsilon_{ij}({\bf{q}})$ is zero
outside this zone.

If $\epsilon_{ij}({\bf{q}})$ is smooth near
$\mathbf{q}=0$, it can be expanded in a Taylor series
\begin{eqnarray*}
\epsilon_{ij}({\bf{q}})=\epsilon_{ij}(\mathbf{0})+
\sum_k q_k \partial_k \epsilon_{ij}({\bf{0}})+
\sum_{kl} q_k q_l 
\partial_k \partial_l \epsilon_{ij}({\bf{0}})+
\cdots
\end{eqnarray*}

The first term corresponds to the electric dipole
approximation,
the second term describes optical activity, the following
term corresponds to electric quadrupole and
magnetic dipole transitions.
When only the first term is kept, then 
Eq.(\ref{epsder}) becomes
$\bar\epsilon_{ij}({\bf{r}}-{\bf{r}}')=
\epsilon_{ij}(\mathbf{0})\delta({\bf{r}}-{\bf{r}}')$
and the constitutive relation is now local
$\mathbf{D}(\mathbf{r})=\epsilon(\mathbf{0})\cdot
\mathbf{E}(\mathbf{r})$.

\subsection{Causality}
A susceptibility is causal if it
satisfies the Kramers-Kronig relations,
that we write in the following form \cite{Jackson}
\begin{eqnarray}
\tilde\chi(\omega)=\frac{1}{2\pi i}
\int_{-\infty}^{+\infty}\frac{\tilde\chi(\omega')}
{\omega'-\omega-i\epsilon}d\omega'. \label{KK1}
\end{eqnarray}

Because of the structure (\ref{chi}), the microscopic
susceptibility $\chi(\mathbf{r},\mathbf{r}';\omega)$ is causal.
Thus, by linearity, its two-scale transform 
$\tilde\chi(\rho,\rho';\mathbf{q},\omega)$ is causal.
To be valid, the macroscopic susceptibility
(\ref{macro5}) has also to be causal.

Starting from Eq.(\ref{KK1}), it can be shown by 
recurrence that, for any integer $n>0$,
\begin{eqnarray}
\tilde\chi^n(\omega)=\frac{1}{2\pi i}
\int_{-\infty}^{+\infty}\frac{\tilde\chi^n(\omega')}
{\omega'-\omega-i\epsilon}d\omega'.\label{KKn}
\end{eqnarray}

If we write Eq.(\ref{macro4bis}) in simplified notation as
\begin{eqnarray}
\bar{\chi}(\omega)=
\langle\tilde\chi(\omega)\rangle+\frac{1}{|C|}\langle
\tilde\chi(\omega)\cdot\nabla G^{\#}\nabla\cdot\tilde\chi(\omega)
\rangle +\cdots,\label{KK}
\end{eqnarray}
we can apply the Kramers-Kronig transform to the
right-hand side of Eq.(\ref{KK}). Then, Eq.(\ref{KKn})
shows that each term of the right-hand side is transformed
into itself by the Kramers-Kronig transform. Therefore
\begin{eqnarray*}
\bar\chi(\omega)=\frac{1}{2\pi i}
\int_{-\infty}^{+\infty}\frac{\bar\chi(\omega')}
{\omega'-\omega-i\epsilon}d\omega'
\end{eqnarray*}
and the macroscopic constitutive relation is causal.

Dolgov and coll. \cite{Dolgov} have pointed out that
$\bar\chi(\mathbf{q},\omega)$ might be non-causal
for $\omega=0$.

\subsection{Symmetry}
To assert the validity of our result,
it is important to check that the symmetry group of
the macroscopic dielectric
constant is the point group of the
crystal space group.

Because of the periodicity of 
$\chi(\mathbf{r},\mathbf{r}')$, the lattice 
translations do not intervene. The other symmetry
operations of the crystal are $T=D+t$, where
$D$ is a (possibly improper) rotation and $t$ is
a translation shorter than the lattice vectors. 
Let $T$ be a crystal symmetry operation,
there is a basis of eigenstates 
$\Psi_n(\mathbf{r})$ such that
\begin{eqnarray*}
\Psi_n(\mathbf{r})
&=&T\Psi_n(\mathbf{r})
=\Psi_n(T^{-1}\mathbf{r}).
\end{eqnarray*}
Using this property we obtain the following transformation
rule for $\chi(\mathbf{r},\mathbf{r}')$ 
\cite{Hanke2,Johnson}
\begin{eqnarray}
\chi_{ij}(\mathbf{r},\mathbf{r}')=\sum_{i'j'} D_{ii'}D_{jj'}
\chi_{i'j'}(T^{-1}\mathbf{r},T^{-1}\mathbf{r}').\label{symet1}
\end{eqnarray}
The presence of the matrices $D_{ii'}$ comes from the derivatives
in definition (\ref{velocity}), which get rid of the translation part
of $T$.

Using Eq.(\ref{symet1}) in Eq.(\ref{twoscalechi}), we find
\begin{eqnarray*}
\tilde{\chi}_{ij}({\bf{\rho}},{\bf{\rho}}'; {\bf{q}}) &=& 
|C|\sum_{s}
\exp[-i{\bf{q}}\cdot({\bf{\rho}}-{\bf{\rho}}'+{\bf{R}}_s)]
\\&&\sum_{i'j'}D_{ii'}D_{jj'}
\chi_{i'j'}(T^{-1}({\bf{\rho}}+{\bf{R}}_s),T^{-1}{\bf{\rho}}').
\end{eqnarray*}

We make the transformation $T^{-1}({\bf{\rho}}+{\bf{R}}_s)=
D^{-1}({\bf{\rho}}+{\bf{R}}_s)-t
=T^{-1}{\bf{\rho}}+D^{-1}{\bf{R}}_s$, we write the argument
of the exponential function as
${\bf{q}}\cdot({\bf{\rho}}-{\bf{\rho}}'+{\bf{R}}_s)=
D{\bf{q}}\cdot(T^{-1}{\bf{\rho}}-T^{-1}{\bf{\rho}}'+D^{-1}{\bf{R}}_s)$,
we use the fact that, for any space group, $D$ is a symmetry
operation of the Bravais lattice, so that $D^{-1}{\bf{R}}_s$
is a lattice vector and we obtain
\begin{eqnarray}
\tilde{\chi}_{ij}({\bf{\rho}},{\bf{\rho}}'; {\bf{q}}) &=&
\sum_{i'j'}D_{ii'}D_{jj'}
\tilde{\chi}_{i'j'}(T^{-1}{\bf{\rho}},T^{-1}{\bf{\rho}}';D{\bf{q}}).
\label{symet2}
\end{eqnarray}

The final step is to replace all the $\tilde\chi$ in the
right-hand side of Eq.(\ref{macro4bis}) by the right-hand side
of Eq.(\ref{symet2}).  For all integrals we make the 
change of variable $\underline{\rho}=T^{-1}{\bf{\rho}}$,
then we use the
fact that $G^{\#}(T\underline{\rho}-T\underline{\rho}')=
G^{\#}(\underline{\rho}-\underline{\rho}')$, the identity
$\partial_{\rho_i}/\partial_{\underline{\rho}_j}=D_{ij}$, 
the orthogonality of
$D$ and the periodicity of all functions involved to prove that
\begin{eqnarray}
\bar{\chi}_{ij}({\bf{q}}) &=& \sum_{i'j'}D_{ii'}D_{jj'}
\bar{\chi}_{i'j'}(D{\bf{q}}).\label{chisym}
\end{eqnarray}

For the dielectric constant $\bar{\epsilon}_{ij}({\bf{0}})$,
we restrict Eq.(\ref{chisym}) to the electric dipole
contribution $\mathbf{q}=\mathbf{0}$ and
we obtain the expected result
\begin{eqnarray*}
\bar{\epsilon}_{ij}({\bf{0}}) &=&
\sum_{i'j'}D_{ii'}D_{jj'}
\bar{\epsilon}_{i'j'}({\bf{0}}).
\end{eqnarray*}

\subsection{Energy density}
The microscopic energy balance is \cite{Duffin}:
\begin{eqnarray*}
-\int d\mathbf{r}\mathbf{j}(\mathbf{r},t)\cdot
\mathbf{e}(\mathbf{r},t)&=&
\frac{1}{\mu_0}\int d\mathbf{r}\mathbf{b}(\mathbf{r},t)\cdot
\frac{\partial\mathbf{b}(\mathbf{r},t)}{\partial t}
\\&&\hspace*{-33mm}+
\epsilon_0
\int d\mathbf{r}\mathbf{e}(\mathbf{r},t)\cdot
\frac{\partial\mathbf{e}(\mathbf{r},t)}{\partial t}+
\frac{1}{\mu_0}\int_\Sigma \mathbf{e}(\mathbf{r},t)\times
\mathbf{b}(\mathbf{r},t)\cdot \mathbf{d}\sigma.\label{energie}
\end{eqnarray*}
The definition of $\mathbf{d}(\mathbf{r})$ gives us
\begin{eqnarray*}
\epsilon_0\frac{\partial\mathbf{e}(\mathbf{r},t)}{\partial t}=
\frac{\partial\mathbf{d}(\mathbf{r},t)}{\partial t}-
\mathbf{j}(\mathbf{r},t),
\end{eqnarray*}
and the energy balance can be written
\begin{eqnarray*}
\int d\mathbf{r}u(\mathbf{r},t)+
\int_\Sigma \mathbf{s}(\mathbf{r},t)
\cdot \mathbf{d}\sigma=0,
\end{eqnarray*}
where the microscopic electromagnetic energy is
\begin{eqnarray*}
u(\mathbf{r},t)=
\frac{1}{\mu_0}\int d\mathbf{r}\mathbf{b}(\mathbf{r},t)\cdot
\frac{\partial\mathbf{b}(\mathbf{r},t)}{\partial t}+
\int d\mathbf{r}\mathbf{e}(\mathbf{r},t)\cdot
\frac{\partial\mathbf{d}(\mathbf{r},t)}{\partial t},
\end{eqnarray*}
and the microscopic Poynting vector is
\begin{eqnarray*}
\mathbf{s}(\mathbf{r},t)=
\frac{1}{\mu_0}\mathbf{e}(\mathbf{r},t)\times
\mathbf{b}(\mathbf{r},t).
\end{eqnarray*}

If we restrict the definition of the macroscopic
fields to the average of the first term in expansion
(\ref{asumtype}), we want to determine whether
the average of the microscopic Poynting vector is
equal to the macroscopic Poynting vector
$\mathbf{S}(\mathbf{r},t)=
\frac{1}{\mu_0}\mathbf{E}(\mathbf{r},t)\times
\mathbf{B}(\mathbf{r},t)$, and whether the
average of the microscopic energy density is
equal to the macroscopic energy density.
If this were not the case, the energy
arguments using the macroscopic Maxwell equations would lack
any microscopic basis.

For the Poynting vector, the answer is immediately
yes, because we have shown that
${\mathbf{b}^{(0)}}(\mathbf{q},\mathbf{\rho})=
\mathbf{B}(\mathbf{q})$. Therefore
${\mathbf{b}^{(0)}}(\mathbf{r},t)=\mathbf{B}(\mathbf{r},t)$,
the magnetic field does not oscillate rapidly, it
is equal to its average and
\begin{eqnarray*}
\langle\mathbf{s}(\mathbf{r},t)\rangle &=&
\frac{1}{\mu_0}\langle\mathbf{e}^{(0)}(\mathbf{r},t)\times
\mathbf{b}^{(0)}(\mathbf{r},t)\rangle\\
&=&
\frac{1}{\mu_0}\langle\mathbf{e}^{(0)}(\mathbf{r},t)\rangle\times
\mathbf{B}(\mathbf{r},t)
=
\frac{1}{\mu_0}\mathbf{E}(\mathbf{r},t)\times
\mathbf{B}(\mathbf{r},t)\\
&=&
\mathbf{S}(\mathbf{r},t).
\end{eqnarray*}

The proof is similar for the magnetic part of the
energy density. However, the problem is more difficult
for the electric part, because it is a product
of two functions that oscillate very rapidly,
and it is not obvious that the average of the product
is equal to the product of the averages.
We prove now that the average of the
microscopic energy density is indeed given by the
usual macroscopic formula.

First, we Fourier transform the fields in space and
time as
\begin{eqnarray*}
\mathbf{e}^{(0)}(\mathbf{r},t)=\int_{BZ} d\mathbf{q}
\int d\omega \exp[i(\mathbf{q}\cdot\mathbf{r}-\omega t)]
\mathbf{e}^{(0)}(\mathbf{q},\mathbf{r})
\end{eqnarray*}
(the variable $\omega$ is still implicit for the fields).

We need to show that
$\langle
\mathbf{e}^{(0)}(\mathbf{q}',\mathbf{\rho})\cdot
\mathbf{d}^{(0)}(\mathbf{q},\mathbf{\rho})\rangle$
is equal to the corresponding macroscopic product
$\mathbf{E}(\mathbf{q}')\cdot\mathbf{D}(\mathbf{q})$.

The microscopic displacement field
$\mathbf{d}^{(0)}(\mathbf{r},\mathbf{q})$ is given
by Eq.(\ref{macro2}) and the eletric field by
${\bf{e}}^{(0)}({\bf{q}},{\bf{\rho}})=
{\bf{E}}({\bf{q}})-\nabla_\rho\phi({\bf{q}},{\bf{\rho}})$.
The macroscopic displacement field is
given by Eq.(\ref{macro3}), so that
\begin{eqnarray*}
\langle
\mathbf{e}^{(0)}(\mathbf{q}',\mathbf{\rho})\cdot
\mathbf{d}^{(0)}(\mathbf{q},\mathbf{\rho})\rangle
&=&\mathbf{E}(\mathbf{q}')\cdot
\mathbf{D}(\mathbf{q})
\\&&\hspace*{-30mm}-
\langle \nabla_\rho\phi(\mathbf{q}',\rho)\cdot
\tilde\chi(\rho,\rho';\mathbf{q})\rangle_{\rho\rho'}
\cdot\mathbf{E}(\mathbf{q})
\\&&\hspace*{-30mm}+
\epsilon_0\langle \nabla_\rho\phi(\mathbf{q}',\rho)\cdot
\nabla_\rho\phi(\mathbf{q},\rho)\rangle
\\&&\hspace*{-30mm}+
\langle \nabla_\rho\phi(\mathbf{q}',\rho)\cdot
\tilde\chi(\rho,\rho';\mathbf{q})
\cdot\nabla_{\rho'}\phi(\mathbf{q},\rho')\rangle_{\rho\rho'},
\end{eqnarray*}
where we have used the fact that 
$\langle\nabla_\rho\phi(\mathbf{q},\rho)\rangle=0$.
Integrating by parts to eliminate the gradient of
$\phi(\mathbf{q}',\rho)$ we obtain
\begin{eqnarray*}
\langle
{\mathbf{e}^{(0)}}(\mathbf{q}',\mathbf{\rho})\cdot
\mathbf{d}^{(0)}(\mathbf{q},\mathbf{\rho})\rangle
&=&\mathbf{E}(\mathbf{q}')\cdot
\mathbf{D}(\mathbf{q})
\\&&\hspace*{-30mm}+
\langle \phi(\mathbf{q}',\rho)[
\nabla_\rho\cdot\tilde\chi(\rho,\rho';\mathbf{q})
\cdot\mathbf{E}(\mathbf{q})
\\&&\hspace*{-30mm}-
\epsilon_0 \Delta_\rho\phi(\mathbf{q},\rho)-
\nabla_\rho\cdot
\tilde\chi(\rho,\rho';\mathbf{q})
\cdot\nabla_{\rho'}\phi(\mathbf{q},\rho')]\rangle_{\rho\rho'}\\
&=&\mathbf{E}(\mathbf{q}')\cdot
\mathbf{D}(\mathbf{q}),
\end{eqnarray*}
where the last step was derived using the cell equation (\ref{cell}).

A mathematical study of the convergence
of the energy density was carried out for a local 
susceptibility by Markowich and Poupaud \cite{Markowich}.

When magnetic properties are taken into account,
that simple result could fail. If this were the
case, we would have a microscopic basis for the
non-standard Poynting vectors investigated in
Ref.\cite{Rikken}.

\subsection{Reaction field matrix}
It is also possible to integrate by parts to apply
both gradients to the Green function, but 
the double gradient of the Green function has a
singularity that must be treated with care
\cite{Frahm,Weiglhofer}. On the other hand, this
form has the advantage of recovering the usual
dipole-dipole interaction of classical dielectric
theory \cite{Scaife} as will be shown now.
In Eq.(\ref{Meq}), an integration by parts
transfers the derivative $\partial_{\rho'_j}$
from $f^{n'}_j(\mathbf{\rho}',\mathbf{q})$ to
$G^{\#}(\rho-\rho')$, then 
$\partial_{\rho'_j}G^{\#}(\rho-\rho')=-\partial_{\rho_j}G^{\#}(\rho-\rho')$.
Therefore, if we define the matrix Green function
$\mathbf{G}^{\#}(\rho-\rho')$ by
\begin{eqnarray*}
G^{\#}_{ij}(\rho-\rho')=
\partial_{\rho_i}\partial_{\rho_j}G^{\#}(\rho-\rho'),
\end{eqnarray*}
The definition of the
reaction field matrix becomes
\begin{eqnarray*}
M({\bf{q}})_{nn'}&=&|C|\sum_{ij}
\langle \mathbf{g}^n({\bf{\rho}},{\bf{q}})\cdot
\mathbf{G}^{\#}(\rho-\rho')\cdot
\mathbf{f}^{n'} ({\bf{\rho}}',{\bf{q}})\rangle_{\rho\rho'}.
\end{eqnarray*}

To proceed, we write 
\begin{eqnarray*}
\mathbf{G}^{\#}(\rho-\rho')=\mathbf{G}^0(\rho-\rho')+
\sum_{s\not= 0}\mathbf{G}^0(\rho-\rho'+\mathbf{R}_s).
\end{eqnarray*}

Weiglhofer has shown that \cite{Frahm,Weiglhofer}
\begin{eqnarray*}
G_{ij}^0(\mathbf{r})&=&\frac{1}{4\pi\epsilon_0}
\partial_{r_i}\partial_{r_j} \frac{1}{r}\\
&=&-\frac{1}{3\epsilon_0}\delta(\mathbf{r})\delta_{ij}
+\frac{1}{4\pi\epsilon_0}\frac{3r_ir_j-r^2\delta_{ij}}{r^5}.
\end{eqnarray*}
The first term gives the depolarization term of the
classical Lorentz theory \cite{Jackson}, the second
one is more delicate and is not considered in the
classical approach.

Finally, we shall need the multiple-scattering expression
for $G^0(\rho-\rho'+\mathbf{R}_s)$, valid when
$|\rho|+|\rho'|<|\mathbf{R}_s|$ 
(see Refs.\cite{Sack,Schadler} for this and the most
general cases).

\begin{eqnarray*}
G_0(\mathbf{\rho}-\mathbf{\rho}'+\mathbf{R}_s)&=&\sum_{\ell m\ell'm'}
Y_\ell^m(\hat\mathbf{\rho})j_\ell(\rho)H_{\ell m\ell'm'}(\mathbf{R}_s)
\\&&
{Y_{\ell'}^{m'}}^*(\hat\mathbf{\rho}')j_{\ell'}(\rho')
\end{eqnarray*}
where $j_{\ell}(\rho)=\rho^\ell/(2\ell+1)!!$ and
\begin{eqnarray}
H_{\ell m\ell'm'}(\mathbf{R}_s)&=&-4\pi (-1)^\ell 
C_{\ell m \ell+\ell' m'-m}^{\ell'm'}
Y_{\ell+\ell'}^{m'-m}(\hat\mathbf{R}_s)\nonumber\\&&
\frac{(2\ell+2\ell'-1)!!}{R_s^{\ell+\ell'+1}}, \label{MST}
\end{eqnarray}
with $(-1)!!=1$ and where $C_{\ell m \ell+\ell' m'-m}^{\ell'm'}$
is a Gaunt coefficient.

Because of the derivatives in the definition of
$\mathbf{G}^{\#}(\rho-\rho')$, the terms $\ell=0$ 
or $\ell'$=0 in
Eq.(\ref{MST}) do not contribute. Therefore, the
cell-to-cell electrostatic interaction begins with
a $1/R_s^3$. In other words, there is no interaction
between the cell total electron charges
(that would be in $1/R_s$),
although the cell are not neutral. To be more
explicit, the cells are neutral if one consideres
the nuclei, but the nuclei charges did not enter
our description, and their contribution to the
dielectric function is very small at optical energies.
Therefore, there is no $1/R_s$ interaction although
only electrons are taken into account.

For an octahedral or a tetrahedral symmetry, group theory
tells us that $\sum_s
Y_{\ell}^{m}(\hat\mathbf{R}_s)$ is zero for
$\ell$=1,2 and 3. Therefore, the electrostatic interaction
starts with a $1/R_s^5$ dependence.

\section{Bloch wave functions}

Many works are devoted to the calculation of the
optical response of semiconductors, using a band-structure
approach. To deal with this case, we specify now
our results to the case of one-electron wavefunctions.

The Bloch theorem tells us that one-electron wavefunctions
in a crystal can be written 
$\exp(i\mathbf{k}\cdot\mathbf{r})u_l(\mathbf{r};\mathbf{k})$,
where the index $l$ runs over the bands.
When this equation is introduced into the velocity
matrix elements (\ref{velocity}), we find
\begin{eqnarray*}
\mathbf{V}^{ll'}(\mathbf{r};\mathbf{k},\mathbf{k}')&=&
-\frac{i\hbar}{2m}
\exp[i(\mathbf{k}'-\mathbf{k})\cdot\mathbf{r}]\\&&
(n_{l',\mathbf{k}'}-n_{l,\mathbf{k}})
\mathbf{W}^{ll'}(\mathbf{r};\mathbf{k},\mathbf{k}'),
\end{eqnarray*}
where $n_{l,\mathbf{k}}$ is the occupation of 
the $l$-th band with Bloch vector $\mathbf{k}$
and the periodic functions
$\mathbf{W}^{ll'}(\mathbf{r};\mathbf{k},\mathbf{k}')$ are
\begin{eqnarray*}
\mathbf{W}^{ll'}(\mathbf{r};\mathbf{k},\mathbf{k}')&=&
u_l^*(\mathbf{r};\mathbf{k})\nabla
u_{l'}(\mathbf{r};\mathbf{k}')-
u_{l'}(\mathbf{r};\mathbf{k}')\nabla
u_l^*(\mathbf{r};\mathbf{k})\\&&
+i(\mathbf{k}+\mathbf{k}')
u_l^*(\mathbf{r};\mathbf{k})
u_{l'}(\mathbf{r};\mathbf{k}').
\end{eqnarray*}
Neglecting the first and the last term in Eq.(\ref{chi})
we find
\begin{eqnarray*}
\chi_{ij}(\mathbf{r},\mathbf{r}')&=&-\frac{e^2\hbar^2}{4\omega^2 m^2}
\int d\mathbf{k} d\mathbf{k}'
\exp[i(\mathbf{k}'-\mathbf{k})\cdot(\mathbf{r}-\mathbf{r}')]
\\&&\sum_{ll'}
(n_{l',\mathbf{k}'}-n_{l,\mathbf{k}})
\frac{W_i^{ll'}(\mathbf{r};\mathbf{k},\mathbf{k}')
W_j^{l'l}(\mathbf{r};\mathbf{k}',\mathbf{k})}
{E_{l'}(\mathbf{k}')-E_{l}(\mathbf{k})-\hbar\omega}.
\end{eqnarray*}

We treat now the electric dipole case $\mathbf{q}=0$.
From the definition of the two-scale transform of $\chi$ and
from identity (\ref{somsurs}) we find
\begin{eqnarray}
\tilde\chi_{ij}(\mathbf{\rho},\mathbf{\rho}';\mathbf{0})&=&
-\frac{(2\pi)^3 e^2\hbar^2}{4\omega^2 m^2}
\int d\mathbf{k} \sum_{ll'}
(n_{l',\mathbf{k}}-n_{l,\mathbf{k}})\\&&
\frac{W_i^{ll'}(\mathbf{\rho};\mathbf{k},\mathbf{k})
W_j^{l'l}(\mathbf{\rho}';\mathbf{k},\mathbf{k})}
{E_{l'}(\mathbf{k})-E_{l}(\mathbf{k})-\hbar\omega}.
\label{chiband}
\end{eqnarray}

Therefore, in the electric dipole
case, the transitions are vertical.
When spatial dispersion is investigated, then the transitions
are no longer vertical, they explore a part of the
bands around the vertical \cite{Koopmans}.

We can use the
Schr\"odinger equation for $u_{l}$ and $u_{l'}$ to show
that
\begin{eqnarray}
\nabla_\rho\cdot
\mathbf{W}^{ll'}(\mathbf{\rho};\mathbf{k},\mathbf{k})&=&
-\frac{2m}{\hbar^2}(E_{l'}(\mathbf{k})-E_{l}(\mathbf{k}))
u_l^*(\mathbf{\rho};\mathbf{k})
u_{l'}(\mathbf{\rho};\mathbf{k}).
\nonumber\\&&\label{densiden}
\end{eqnarray}
In Ref.\cite{Keller}, p.121, Keller has proved that 
such a relation between 
$\nabla\cdot\mathbf{V}^{0n}(\mathbf{r})$ and matrix elements
of the density operator holds
also valid in the many-body case.

To use the macroscopic constitutive relation (\ref{macro5})
we must define a separable form (\ref{sepchi})
for $\tilde\chi$ in Eq.(\ref{chiband}). We can choose
\begin{eqnarray*}
f_i^{ll'}(\rho;\mathbf{k})&=&-\frac{(2\pi)^3 e^2\hbar^2}{4\omega^2 m^2}
(n_{l',\mathbf{k}}-n_{l,\mathbf{k}})\\&&
\frac{W_i^{ll'}(\mathbf{\rho};\mathbf{k},\mathbf{k})}
{E_{l'}(\mathbf{k})-E_{l}(\mathbf{k})-\hbar\omega},
\nonumber\\
g_j^{\underline{l}\,\underline{l}'}
(\rho';\underline{\mathbf{k}})&=&
W_j^{\underline{l}'\,\underline{l}}
(\mathbf{\rho}';\underline{\mathbf{k}},\underline{\mathbf{k}}).
\end{eqnarray*}
Finally, using identity (\ref{densiden}) and the
alternative definition of the reaction field matrix
$M$ we find
\begin{eqnarray*}
M^{ll'\underline{l}\,\underline{l}'}_{\mathbf{k}\underline{\mathbf{k}}}
&=&\frac{|C| (2\pi)^3 e^2}{(\hbar\omega)^2}
(n_{l',\mathbf{k}}-n_{l,\mathbf{k}})\\&&
\frac{(E_{l'}(\mathbf{k})-E_{l}(\mathbf{k}))
(E_{\underline{l}}(\underline{\mathbf{k}})-
E_{\underline{l}'}(\underline{\mathbf{k}}))}
{(E_{l'}(\mathbf{k})-E_{l}(\mathbf{k})-\hbar\omega)}
\\&&
\hspace*{-9mm}
\langle u_l^*(\mathbf{\rho};\mathbf{k})
u_{l'}(\mathbf{\rho};\mathbf{k})
G^{\#}(\rho-\rho')
u_{\underline{l}'}^*(\mathbf{\rho}';\underline{\mathbf{k}})
u_{\underline{l}}(\mathbf{\rho}';\underline{\mathbf{k}})
\rangle_{\rho\rho'}.
\end{eqnarray*}
Because of the presence of the Bloch vector indices
$\mathbf{k}$ and $\underline{\mathbf{k}}$, the reaction
field matrix is huge and its inversion will be a heavy
computation.

All the ingredients are now given for a
band-structure calculation of the macroscopic dielectric
constant in the electric dipole approximation. 
The final formula is 
\begin{eqnarray*}
\bar\epsilon_{ij}&=&\epsilon_0\delta_{ij}-
\frac{(2\pi)^3 e^2\hbar^2}{4m^2\omega^2}
\int d\mathbf{k} d\underline{\mathbf{k}}
\sum_{ll'\underline{l}\,\underline{l}'}
(n_{l',\mathbf{k}}-n_{l,\mathbf{k}})\\&&
\frac{\langle W_i^{ll'}(\mathbf{\rho};\mathbf{k},\mathbf{k})\rangle
{[(1-M)^{-1}]}^
{ll'\underline{l}\,\underline{l}'}_{\mathbf{k}\underline{\mathbf{k}}}
\langle W_j^{\underline{l}'\underline{l}}
(\mathbf{\rho};\underline{\mathbf{k}},\underline{\mathbf{k}})\rangle}
{(E_{l'}(\mathbf{k})-E_{l}(\mathbf{k})-\hbar\omega)}
\end{eqnarray*}

A similar formula can be obtained for spatial dispersion 
($\mathbf{q}\not=\mathbf{0}$), adapting the 
implementation described in Ref.\cite{Rezvani},
where the Coulomb singularity and the Umklapp
processes are treated in detail.

\section{Many-body dielectric function}

Before calculating the dielectric constant corresponding
to a many-body susceptibility, we must show that the
transverse part of the microscopic electric field varies
slowly.
All fields can be written as the sum of a transverse and
a longitudinal components. The transverse component of 
$\mathbf{e}(\mathbf{q},\mathbf{r})$ is defined by
$\nabla\cdot\mathbf{e}_T(\mathbf{r})=0$ and
$\nabla\times\mathbf{e}_T(\mathbf{r})=
\nabla\times\mathbf{e}(\mathbf{r})$.
The corresponding equations for the two-scale transforms are
\begin{eqnarray*}
i{\bf{q}}\cdot\mathbf{e}_T(\mathbf{q},\mathbf{\rho}) +
\nabla_\rho\cdot\mathbf{e}_T(\mathbf{q},\mathbf{\rho})&=&0\\
(i{\bf{q}}+\nabla_\rho)
\times\mathbf{e}_T(\mathbf{q},\mathbf{\rho})&=&
(i{\bf{q}}+\nabla_\rho)
\times\mathbf{e}(\mathbf{q},\mathbf{\rho}).
\end{eqnarray*}

Using the expansion (\ref{asumtype}) for the full and 
transverse electric fields, we obtain for the $a^{-1}$ term:
\begin{eqnarray}
\nabla_\rho\cdot\mathbf{e}^{(0)}_T(\mathbf{q},\mathbf{\rho})&=&0
\nonumber\\
\nabla_\rho\times\mathbf{e}^{(0)}_T(\mathbf{q},\mathbf{\rho})&=&
\nabla_\rho\times\mathbf{e}^{(0)}(\mathbf{q},\mathbf{\rho})=0, \label{eqet}
\end{eqnarray}
where the last equality was obtained from the first of 
Eqs.(\ref{drhoq}). Equations (\ref{eqet})
are the same as the equations for 
$\mathbf{b}^{(0)}(\mathbf{q},\mathbf{\rho})$ in 
section 4. Therefore, the conclusion is the same
and $\mathbf{e}^{(0)}_T(\mathbf{q},\mathbf{\rho})=
\mathbf{F}(\mathbf{q})$ does not depend on $\rho$,
it is a slowly varying function.

To determine $\mathbf{F}(\mathbf{q})$ we
use the following equation given by the $a^{0}$ term
\begin{eqnarray*}
i{\bf{q}}\cdot\mathbf{e}^{(0)}_T(\mathbf{q},\mathbf{\rho}) +
a\nabla_\rho\cdot\mathbf{e}^{(1)}_T(\mathbf{q},\mathbf{\rho})&=&0.\
\end{eqnarray*}
The average of both sides yields
$\mathbf{q}\cdot\mathbf{F}(\mathbf{q})=0$ and $\mathbf{F}(\mathbf{q})$
is a macroscopically transverse vector.
The other equation given by the $a^{0}$ term is
\begin{eqnarray*}
i{\bf{q}}\times\mathbf{e}^{(0)}_T(\mathbf{q},\mathbf{\rho}) +
\nabla_\rho\times\mathbf{e}^{(1)}_T(\mathbf{q},\mathbf{\rho})&=&
i{\bf{q}}\times\mathbf{e}^{(0)}(\mathbf{q},\mathbf{\rho}) \\&&+
\nabla_\rho\times\mathbf{e}^{(1)}(\mathbf{q},\mathbf{\rho}).
\end{eqnarray*}
Taking again the average of both sides we obtain
$\mathbf{q}\times\mathbf{F}(\mathbf{q})=
\mathbf{q}\times\mathbf{E}(\mathbf{q})$, so that
$\mathbf{F}(\mathbf{q})=\mathbf{E}_T(\mathbf{q})$,
and the microscopic transverse electric field is equal to the
macroscopic transverse electric field.

Notice that this conclusion follows from the Maxwell equations
and does not depend on the constitutive relations.

In many-body theories,
\cite{Keller,Martin,Falk,Bagchi,Falter}
the induced current is related to the 
external electric field or to the
{\em transverse part} of the total
electric field (see Ref.\cite{Keller} for a detailed derivation).
Since the experimental dielectric function relates the macroscopic
electric displacement to the macroscopic total electric field,
we choose the latter formulation and write
\begin{eqnarray*}
\mathbf{d}({\bf{r}})=\epsilon_0\mathbf{e}({\bf{r}})+
\int d{\bf{r}}'
\alpha({\bf{r}},{\bf{r}}')\cdot\mathbf{e}_T({\bf{r}}').
\end{eqnarray*}

Repeating all the steps of section 6, we arrive at 
\begin{eqnarray*}
\mathbf{d}({\bf{q}},{\bf{\rho}})=
\epsilon_0 \mathbf{e}({\bf{q}},{\bf{\rho}})
+\langle
\tilde\alpha({\bf{\rho}},{\bf{\rho}}';{\bf{q}})\cdot
\mathbf{e}_T({\bf{q}},{\bf{\rho}}')\rangle_{\rho'}.
\end{eqnarray*}
Since $\mathbf{e}_T({\bf{q}},{\bf{\rho}}')=
\mathbf{E}_T(\mathbf{q})$ does not depend on $\rho'$,
the average yields simply
\begin{eqnarray*}
\mathbf{D}({\bf{q}})=
\epsilon_0 \mathbf{E}({\bf{q}})
+\langle
\tilde\alpha({\bf{\rho}},{\bf{\rho}}';{\bf{q}})\rangle_{\rho\rho'}\cdot
\mathbf{E}_T({\bf{q}}).
\end{eqnarray*}

$\mathbf{E}_T({\bf{q}})$ can be expressed in terms of 
$\mathbf{E}({\bf{q}})$ by
\begin{eqnarray*}
\mathbf{E}_T(\mathbf{q})=\mathbf{E}(\mathbf{q})-
\frac{\mathbf{q}\cdot\mathbf{E}(\mathbf{q})}{q^2}\mathbf{q}
\end{eqnarray*}
and the final formula for the many-body case is
$\mathbf{D}(\mathbf{q})=
\epsilon(\mathbf{q})\cdot\mathbf{E}(\mathbf{q})$ with
\begin{eqnarray*}
\epsilon_{ij}(\mathbf{q})=
\epsilon_0 \delta_{ij}
+\sum_{k}\langle
\tilde\alpha_{ik}({\bf{\rho}},{\bf{\rho}}';{\bf{q}})\rangle_{\rho\rho'}
\frac{q^2\delta_{kj}-q_kq_j}{q^2}.
\end{eqnarray*}

This formula looks simpler than for the case of the
random phase approximation, but the calculation of the
many-body susceptibility 
$\alpha({\bf{r}},{\bf{r}}')$ is much more difficult, since
it must account for all electron-electron interactions,
including the reaction fields. Notice also that the
limit $\mathbf{q}\rightarrow 0$ is ambiguous because
constant electric field cannot be uniquely decomposed
into a transverse and a longitudinal parts.

\section{Conclusion}

Homogenization theory is usually used
to calculate the properties of ``real'' materials
(porous, fibrous, disordered, etc.). It is a somewhat
just reward that applied physics can also be useful to
basic physics.

Here, homogenization theory was used to calculate the
macroscopic dielectric constant from the microscopic
dielectric function.
Compared to previous works, our approach
is not restricted to cubic materials and provides new
equations to describe the local field effect in 
dielectrics.
This article does not exhaust
the prospects of homogenization theory in solid-state
physics.  It is rather
a detailed presentation of the simplest possible case
and the present study can be developed
in many directions. 

A complete derivation of the macroscopic Maxwell
equations should also take magnetism into account.
In particular,
the mysterious relation between microscopic
and macroscopic magnetic properties \cite{Hirst}
could be handled with homogenization theory,
as well as the question of the general form
of the constitutive relations in bianisotropic
media \cite{Graham}.

Further terms of the expansion can be
calculated \cite{Friedman} to investigate the case
when the wavelength of the incident wave is not
very large as compared to the unit cell, as in the
VUV range and for some near-field optics or
inelastic scattering experiments.

We have considered an infinite crystal, but
homogenization theory can also treat finite
crystals \cite{Wellander,Nguetseng,Allaire,Conca}. 
This is particularly
interesting when the medium exhibits spatial
dispersion, i.e. $q\not=0$ in Eq.(\ref{macro5}),
because additional boundary conditions can be
required to determine the
waves inside the dielectric body 
\cite{Agranovich,Cho,Graham}. 
Homogenization theory is well suited
to describe the boundary layer that forms at the surface
of the dielectric body, and to derive the corresponding
boundary conditions. To do this, one adds to the
bulk (periodic) functions a boundary function that decreases 
exponentially out of the dielectric \cite{Sanchez2,Bakhalov}.
Besides, the Bloch decomposition of the electromagnetic field 
seems to be a promissing alternative for that purpose
\cite{Oster,Conca,Johnson2}.

The present work was carried out within the linear response
approximation. Homogenization theory is fully developed to
deal with non-linear equations \cite{Jikov,Wellander,Pankov}.

We have considered periodic media, but homogenization
theory applies also when the periodic structure varies slowly
\cite{Bensoussan} or in disordered or polycrystalline
media \cite{Jikov}.

Finally, the method is not restricted to electrodynamics
and can be used to calculate any constitutive relation
corresponding to a microscopic non-local equation.

\section{Aknowledgements}
I am very grateful to Prof. Dr. H. Bross for his
help concerning the many-body dielectric function.
I thank M. Ruiz-L\'opez, A. Bossavit and
O. Keller who sent me reprints of their works.
I thank Ph. Sainctavit for his thorough reading of the
manuscript.


\begin{thebibliography}{10}
\bibitem{Developpement}
The expansion does not always start with an $a^0$ term. For instance, the
  expansion of the charge density is $n(\mathbf{q},\rho)=a^{-1}
  n^{(-1)}(\mathbf{q},\rho)+ a^{0} n^{(0)}(\mathbf{q},\rho)+\cdots$.
\bibitem{intder}
The fact that $\langle \nabla_\rho \psi(\mathbf{q},\rho)\rangle=0$ can be shown
  for instance for $\rho_i=x$. In the integral over the unit cell we isolate
  the integral over $x$ and we use $\int_{-L}^{L} dx \partial_x
  \phi({\bf{q}},x,y,z)= \phi({\bf{q}},L,y,z)-\phi({\bf{q}},-L,y,z)$, which is
  zero by periodicity of $\phi({\bf{q}},{\bf{\rho}})$.
\end{thebibliography}
\end{document}